\documentclass[aps,prd,twocolumn,showpacs,10pt,superscriptaddress,preprintnumbers,nofootinbib,floatfix]{revtex4-1}
\usepackage[utf8]{inputenc}
\usepackage{amsmath,amssymb,bm,slashed,braket}
\usepackage{graphicx}
\usepackage{epstopdf}
\usepackage[dvipsnames,table]{xcolor}
\usepackage[normalem]{ulem}
\usepackage{colortbl}
\usepackage{multirow, booktabs}
\usepackage{makecell}
\usepackage{hhline}
\usepackage{dsfont}
\usepackage{verbatim}
\usepackage[colorlinks=true,
  linkcolor=blue,
  urlcolor=blue,
  citecolor=purple,
  bookmarks=true,
  bookmarksnumbered=true,
  breaklinks=true,
  pdfpagemode=Fullscreen,
pdfstartview=FitBH]{hyperref}

\allowdisplaybreaks[4]
\usepackage[capitalise]{cleveref}
\usepackage{stackengine}
\usepackage{orcidlink}
\usepackage{nicematrix}
\usepackage{ctable}
\usepackage{pifont}
\usepackage{cancel}
\usepackage{tikz}
\usepackage{tkz-euclide}
\usetikzlibrary{backgrounds}
\usetikzlibrary{decorations.pathmorphing}
\usetikzlibrary{arrows.meta}
\tikzset{
  mystyle/.style={line width=1, baseline, scale=0.6, every node/.style={scale=1}},
  v/.style={decorate, draw, decoration={snake, segment length=2.mm, amplitude=0.5mm}},
  f/.style={draw, decoration={markings,mark=at position #1 with {\arrow[]{Latex[length=1.5mm,width=1.5mm]}}},
  postaction={decorate},node contents=#1},
  f/.default=.6,
  fb/.style={draw,decoration={markings,mark=at position #1 with {\arrowreversed[]{Latex[length=1.5mm,width=1.5mm]}}},
  postaction={decorate},node contents=#1},
  fb/.default=.6,
  s/.style={dashed,draw, decoration={markings,mark=at position #1 with {\arrow[]{Latex[length=1.5mm,width=1.5mm]}}},
  postaction={decorate},node contents=#1},
  s/.default=.6,
  sb/.style={dashed,draw,decoration={markings,mark=at position #1 with {\arrowreversed[]{Latex[length=1.5mm,width=1.5mm]}}},
  postaction={decorate},node contents=#1},
  sb/.default=.4,
  snar/.style={dashed,draw,line width =1.25pt},
  cross/.style={cross out, draw=black, minimum size=2*(#1-\pgflinewidth), inner sep=0pt, outer sep=0pt},
}

\def\lsim{\mathrel{\raise.3ex\hbox{$<$\kern-.75em\lower1ex\hbox{$\sim$}}}}
\def\gsim{\mathrel{\raise.3ex\hbox{$>$\kern-.75em\lower1ex\hbox{$\sim$}}}}

\newcommand{\calA}{{\cal A}}

\newcommand{\calF}{{\cal F}}

\newcommand{\kvec}{{\bf k}}

\newcommand{\pvec}{{\bf p}}
\newcommand{\Pvec}{{\bf P}}

\newcommand{\pw}{{\rm pw}}
\newcommand{\vx}{{\rm vx}}

\newcommand{\keV}{{\rm keV}}
\newcommand{\MeV}{{\rm MeV}}

\begin{document}

\title{Backward Compton scattering with three vortex particles}

\author{Yi Liao\,\orcidlink{0000-0002-1009-5483}}
\email{liaoy@m.scnu.edu.cn}
\author{Zhaolong Teng\,\orcidlink{0000-0002-7141-2331}}
\email{tengcl@m.scnu.edu.cn}
\author{Hao-Lin Wang\,\orcidlink{0000-0002-2803-5657}}
\email{whaolin@m.scnu.edu.cn}
\affiliation{State Key Laboratory of Nuclear Physics and
  Technology, Institute of Quantum Matter, South China Normal
University, Guangzhou 510006, China}
\affiliation{Guangdong Basic Research Center of Excellence for
  Structure and Fundamental Interactions of Matter, Guangdong
  Provincial Key Laboratory of Nuclear Science, Guangzhou
510006, China}

\begin{abstract}
  We investigate backward Compton scattering in which an incident vortex (vx) photon collides head-on with a plane-wave (pw) electron and both final-state particles are projected onto vortex states, $\gamma_\vx+e^-_\pw\to\gamma_\vx+e^-_\vx$. We derive analytically the scattering amplitude and differential cross sections in the energy and cone angle of either final particle. Rotational symmetry implies a selection rule associated with the conservation of the total angular momentum along the collision axis. A Bessel-Gaussian wave packet is adopted for the incident photon to provide a physical normalization and regularize the boundary singularities of ideal Bessel states. We present numerical results for a $10~\MeV$ electron colliding with a vortex photon of central energy $1~\MeV$ or $10~\keV$. The final-photon distributions exhibit strong energy--angle correlations, topological-charge-dependent interference fringes, and systematic shifts of their dominant peaks, allowing different topological charge sectors to be enhanced through angular or energy post-selection. For a longitudinally polarized incident electron, the dominant channels favor final photons of matching helicity. For the $1~\MeV$ photon benchmark, the final electron can likewise be produced as a MeV-scale vortex state with a sizable cone angle and a large angular-momentum projection. These results demonstrate that triple-vortex backward Compton scattering offers a potential means of generating and controlling high-energy vortex photons and electrons.
\end{abstract}
\maketitle

\section{Introduction}

A vortex state of a particle is characterized by a non-plane-wave wave function with a helical wavefront and a well-defined projection of the intrinsic orbital angular momentum (OAM) along its propagation axis \cite{Bliokh:2017uvr,Lloyd:2017ipi,Knyazev_2018,Ivanov:2022jzh}. Since recognition in the early 1990s that helically phased light can carry quantized OAM \cite{Allen:1992zz}, vortex states have found broad applications in optics, quantum information, and atomic manipulation \cite{torres2011twisted,andrews2012angular,ALLEN1999291,Molina-Terriza:2007ydx,Padgett:17,Knyazev:2018,Babiker_2019}. The intrinsic OAM degree of freedom encoded in such states has also opened new avenues for research in nuclear and particle physics \cite{Wu:2021trm,Lu:2023wrf,Lu:2025tyx,Ivanov:2019vxe,Ivanov:2020kcy}. Realizing these opportunities, however, requires the generation and control of vortex states at high energies.

Experimentally, vortex photons have been routinely generated in the optical regime using fork holograms, spiral phase plates, and related wavefront-shaping elements, enabling applications in optics \cite{Heckenberg1992,Sueda2004,YaoPadgett2011,Willner2015,Mair2001,He1995}. Vortex radiation has also been demonstrated at shorter wavelengths, extending into the extreme-ultraviolet and X-ray regimes through high-harmonic generation and accelerator-based sources, including synchrotron undulators and free-electron lasers \cite{Peele2002,Bahrdt2013,Gauthier2017,Ribic2017,Lee2019,Ivanov:2022jzh}. More recently, evidence for sub-MeV vortex $\gamma$ photons was reported from an all-optical inverse-Compton scattering experiment \cite{Wei:2025zsv}. Electron vortex beams have been experimentally produced in electron microscopes with kinetic energies up to $300\,\keV$ and OAM projections of hundreds, and even more than $1000\hbar$, per electron \cite{Uchida:2010hbm,Verbeeck:2010ezk,McMorran:2011bql,Bliokh:2017uvr,Lloyd:2017ipi}. Despite these advances, the controlled generation of high-energy vortex photons and electrons remains challenging.

Vortex-particle scattering can exhibit coherence and interference effects with no counterpart in plane-wave observables. If only one incoming particle is prepared in a vortex state while all other external particles are resolved as plane waves, the cross section reduces to an incoherent average over the incoming particle's plane-wave components. Coherent effects arise once two or more external particles are prepared or projected in vortex states \cite{Ivanov:2022jzh}. In Compton scattering, previous studies have addressed the upconversion of incoming vortex photons, the generation of structured X rays from vortex electrons \cite{Jentschura:2010ap,Jentschura:2011ih,Ivanov:2011bv,Seipt:2014bxa}, and the production of vortex photons in intense laser fields \cite{Chen_2018}. Collisions with two incoming vortex particles have also been studied using both Bessel \cite{Ivanov:2012na,Ivanov:2016oue,Karlovets:2016dva,Karlovets:2016jrd,Korchagin:2024nen} and Laguerre-Gaussian states \cite{Zhao:2023cwd,Yang:2026byv}. The production of two correlated outgoing vortex particles in off-axis vortex-plane-wave collisions was established on general kinematic grounds \cite{Ivanov:2011tu}. Process-specific studies include all-vortex nonlinear Compton scattering \cite{Liao:2025skb} and off-axis triple-vortex $e^+e^-\to\gamma\gamma$ scattering \cite{Liao:2026gqh}. A corresponding process-specific analysis of linear backward Compton scattering with an incident vortex photon and a plane-wave electron, however, is still lacking.

Besides probing vortex-dependent scattering effects, Compton scattering offers a possible route to producing high-energy vortex particles. The vortex structure of an incoming photon is preserved in exact backward scattering from a plane-wave electron \cite{Jentschura:2010ap,Jentschura:2011ih} and remains approximately preserved at small scattering angles \cite{Ivanov:2011bv}. Although dedicated detectors for direct projection onto vortex states are not yet available, generalized measurement protocols have been proposed, suggesting that their detection may nevertheless be experimentally feasible \cite{Karlovets:2022evc,Karlovets:2022mhb}. These developments motivate a dedicated study of the simultaneous production and correlated distributions of a vortex photon and a vortex electron in the final state.

In this work, we study backward Compton scattering involving three vortex particles. Specifically, the initial photon and both final-state particles, the scattered photon and electron, are in vortex (vx) states, while the initial electron is described by a plane wave (pw), $\gamma_\vx+e^-_\pw \to \gamma_\vx+e^-_\vx$. We develop an analytical framework for this process and adopt a Bessel-Gaussian (BG) wave packet for the incident vortex photon to provide a physical normalization and regularize the boundary singularities of ideal Bessel states. Numerical results are presented for initial vortex photons with energy of $1\,{\MeV}$ or $10\,{\keV}$ scattering from a relativistic plane-wave electron with energy $10\,{\MeV}$. We investigate in detail the distributions of the final-state vortex photon and electron, as well as the dependence of their opening angles and energies on the corresponding topological charges.
These results may facilitate the generation and manipulation of high-energy vortex photons and electrons through backward Compton scattering.

The paper is organized as follows. In \cref{sec:theoretical_calc}, we derive the cross section for backward Compton scattering with three vortex particles. We then perform numerical analyses and discuss the results in \cref{sec:numerical_res}. Our concluding remarks are presented in \cref{sec:conclusion}. The helicity amplitudes for general plane-wave Compton scattering are collected in \cref{sec:helicity_amp} for the sake of reference.

\section{Theoretical calculations}
\label{sec:theoretical_calc}
Let us consider backward Compton scattering in which a plane-wave electron moving along the $z$ axis collides head-on with a counter-propagating Bessel vortex photon along the $-z$ axis to produce a Bessel vortex photon and electron, both propagating along the $z$ axis:
\begin{align}
  &\gamma(k_1\lambda_1;m_1)+e(p_1h_1)\rightarrow \gamma(k_2\lambda_2;m_2)+e(p_2h_2;m_e).
\end{align}
Here, $k_a$ and $p_a$ ($a=1,2$) denote the four-momentum of a plane-wave particle or component of a vortex particle, each of which is in its helicity state with $\lambda_a=\pm$ and $h_a=\pm$. The four-momenta are denoted as
\begin{subequations}
  \begin{align}
    &k_1^\mu=\omega_1(1,\hat{\kvec}_1),\quad p_1^\mu=(E_1,0,0,p_1),
    \\
    &k_2^\mu=\omega_2(1,\hat{\kvec}_2),\quad p_2^\mu=(E_2,\pvec_2),
  \end{align}
\end{subequations}
where the polar and azimuthal angles of various unit vectors are $\hat{\kvec}_a:(\theta_a,\varphi_a)$ and $\hat{\pvec}_2:(\theta_e,\varphi_e)$.
$m_{a,e}$ are the topological charges of the vortex states, also called orbital helicities, i.e., the projection of the total angular momentum of a vortex particle in its own propagation direction \cite{Ivanov:2011kk}. 
For the final vortex electron and photon moving in the $z$ axis, the polar angle is the open (or cone) angle of the vortex and $m_{e,2}$ are measured with respect to the $z$ axis; but for the initial vortex photon propagating in the $-z$ axis, its open angle is $\vartheta=\pi-\theta_1\in[0,\pi/2)$ and its $m_1$ is measured with respect to the $-z$ axis.

The scattering matrix element for the triple-vortex process is
\begin{align}
  S_{\rm vx}&=\iiint \frac{d^2 \kvec_1^\perp}{(2\pi)^2}\frac{d^2 \kvec_2^\perp}{(2\pi)^2}\frac{d^2 \pvec_2^\perp}{(2\pi)^2}
  \nonumber
  \\
  &\qquad\times a_{\kappa_1,-m_1}(\kvec_1^\perp)  a^*_{\kappa_2 m_2}(\kvec_2^\perp) a^*_{\kappa_e m_e}(\pvec_2^\perp) S_{\rm pw},
  \label{eq:S_vx_def}
\end{align}
where $a_{\kappa m}(\kvec^\perp)$ is the kernel for a Bessel vortex with the topological charge $m$ and the magnitude of the transverse momentum $\kappa$,
\begin{align}
  a_{\kappa m}(\kvec^\perp)=i^{-m} e^{im\varphi_k} \sqrt{\frac{2\pi}{\kappa}} \delta(k^\perp-\kappa),
\end{align}
with $\varphi_k$ being the azimuthal angle of $\kvec^\perp$, and $S_\pw$ is the scattering matrix element for the plane-wave process
\begin{align}
  S_{\rm pw}=i (2\pi)^4 \delta^4(p_1+k_1-p_2-k_2)\calA_{\rm pw},
\end{align}
where $\calA_\pw$ is the standard invariant amplitude. The helicity amplitudes for general momenta are recorded in \cref{sec:helicity_amp}.

While the magnitudes of the transverse momenta in \cref{eq:S_vx_def} are trivially integrated out, the azimuthal $\varphi_{2,e}$ integrals are finished with the help of transverse momentum conservation
\begin{align}
  \nonumber
  &\iint d\varphi_2 d\varphi_e
  ~\delta^2(\kvec_2^\perp+\pvec_2^\perp-\kvec_1^\perp) g(\varphi_2,\varphi_e)
  \\
  =&\frac{1}{2 \Delta}\sum_{\eta=\pm}
  g(\varphi_1-\eta\delta_2,\varphi_1+\eta\delta_e),
  \label{eq:planar_integral}
\end{align}
where $\delta_{2,e}=\angle(\kvec_{2,e}^\perp,\kvec_1^\perp)$ are the inner angles of the triangle $\kvec_2^\perp+\pvec_2^\perp=\kvec_1^\perp$ and $\Delta$ is its area
\begin{align}
  \delta_{2,e}=\arccos{\frac{\kappa_1^2+\kappa^2_{2,e}-\kappa^2_{e,2}}{2 \kappa_1 \kappa_{2,e}}},~
  \Delta=\frac{1}{4}\sqrt{-\lambda(\kappa_1^2,\kappa^2_2,\kappa^2_e)},
\end{align}
with $\kappa_a=\omega_a|\sin\theta_a|$, $\kappa_e=p_2|\sin\theta_e|$, $p_a=|\pvec_a|$, and $\lambda(x,y,z)=x^2+y^2+z^2-2xy-2yz-2zx$. 
The third inner angle, $\delta_1=\angle(\kvec^\perp_2,\kvec^\perp_e)=\arccos[(\kappa_2^2+\kappa_e^2-\kappa_1^2)/(2\kappa_2\kappa_e)]$, will be close to $0$ when $|\kappa_2-\kappa_e|\lsim\kappa_1$ and close to $\pi$ when $\kappa_2+\kappa_e\gsim\kappa_1$.
\cref{eq:S_vx_def} becomes
\begin{widetext}
  \begin{align}
    \nonumber
    S_{\rm vx}&= i\delta(q^0)\delta(q^z)i^{m_1+m_2+m_e}
    \sqrt{\frac{\kappa_1\kappa_2\kappa_e}{2\pi}} \frac{1}{2 \Delta}\sum_{\eta=\pm}e^{i\eta(m_2\delta_2-m_e\delta_e)}\int d\varphi_1 e^{-i(m_1+m_2+m_e)\varphi_1}
    \calA_{\rm pw}(\varphi_1,\varphi_1-\eta\delta_2,\varphi_1+\eta\delta_e),
    \label{eq:S_vx_1}
  \end{align}
\end{widetext}
where azimuthal dependence is explicitly indicated in $\calA_\pw(\varphi_1,\varphi_2,\varphi_e)$ and the conservation of energy and longitudinal momentum gives
\begin{subequations}
  \begin{align}
    &q^0=E_1+\omega_1-E_2-\omega_2=0
    \\
    &q^z=p_1+\omega_1\cos\theta_1-p_2\cos\theta_e-\omega_2\cos\theta_2=0.
  \end{align}
\end{subequations}

The helicity amplitudes $\calA_\pw(\varphi_1,\varphi_2,\varphi_e)$ shown in \cref{sec:helicity_amp} exhibit phase factors $e^{\pm i\varphi_a}$ and $e^{\pm i\varphi_e/2}$ associated with the helicities of the photons and the final electron. For an {\em initial} electron with a positive (negative) helicity, corresponding to $h_1=+$ ($h_1=-$), an overall phase factor $e^{i\varphi_e/2}$ ($e^{-i\varphi_e/2}$) from the {\em final} electron can be extracted so that the remaining phases are $\varphi_a-\varphi_e$ and $\varphi_1-\varphi_2$ which are independent of $\varphi_1$ upon substituting $\varphi_2=\varphi_1-\eta\delta_2$ and $\varphi_e=\varphi_1+\eta\delta_e$. In other words, for an initial electron of helicity $h_1/2$, we have generally
\begin{align}
  \nonumber
  &\calA_{\rm pw}(\varphi_1,\varphi_1-\eta\delta_2,\varphi_1+\eta\delta_e)
  \\
  &\quad=e^{ih_1(\varphi_1+\eta\delta_e)/2}\calA_{\rm pw}(0,-\eta\delta_2,\eta\delta_e).
\end{align}
The $\varphi_1$ integral in \cref{eq:S_vx_1} can now be finished to obtain the conservation law of the total angular momentum in the collision $z$ axis:
\begin{align}
  \label{eq:TAM}
  \frac{1}{2}h_1 -m_1 = m_2+m_e.
\end{align}
This conservation law implies that a transversely polarized electron beam, being a superposition of helicity states, could be used to produce final-state vortex particles in a superposition of two different topological charges.
Note the `wrong' sign of $m_1$ which as an orbital helicity is measured with respect to the $-z$ axis. The scattering matrix element becomes
\begin{align}
  \nonumber
  S_{\rm vx}&= i\delta(q^0)\delta(q^z)i^{h_1/2}\delta_{m_1+m_2+m_e,h_1/2}
   \frac{\sqrt{2\pi\kappa_1\kappa_2\kappa_e}}{2 \Delta}
  \\
  &\quad\times \sum_{\eta=\pm}e^{i\eta[m_2\delta_2+(h_1/2-m_e)\delta_e]}
  \calA_{\rm pw}(0,-\eta\delta_2,\eta\delta_e).
  \label{eq:S_vx_2}
\end{align}

Before we calculate the differential cross section, we have to cope with the normalization issue and the nonintegrable singularity caused by $\Delta^{-2}$ when the three transverse momenta lie on a line. As usual, we normalize a Bessel vortex state in a large cylinder of length $L_z$ and radius $R$ and require that the whole-cylinder integral of its temporal density be unity. This yields the normalization factor $N_\textrm{Be}=[\pi/(2E RL_z)]^{1/2}$ with $E$ being the energy of the state. For final-state particles, such normalization factors cancel out between the amplitude squared and final-state density, so that for our case the phase-space measure is effectively
\begin{align}
  \frac{dp_2^zd\kappa_e}{2E_22\pi}\frac{dk_2^zd\kappa_2}{2\omega_22\pi}.
\end{align}
Normalization of the initial states enters the initial flux. For a vortex-plane-wave collision, we employ an average flux $(v_1+\cos\vartheta)V^{-1}$ \cite{Jentschura:2011ih} or $(1+v_1\cos\vartheta)V^{-1}$ \cite{Ivanov:2011kk} which makes no practical difference since $v_1\approx 1$ for an ultrarelativistic electron. Here $V=\pi R^2L_z$ is the volume of the whole space in which the initial plane-wave electron has the normalization factor $N_\pw=[2E_1V]^{-1/2}$. Squaring $S_\vx$ yields a factor of $2\pi\delta(q^0)2\pi\delta(q^z)\to TL_z$ where $T$ is the duration of interaction. These normalization factors combine in the differential cross section into
$TL_z/T\cdot[\pi/(2\omega_1RL_z)][1/(2E_1V)]\cdot V/(v_1+\cos\vartheta)=1/(2E_12\omega_1)\cdot[1/(v_1+\cos\vartheta)]\cdot[\pi/R]$, leaving behind a factor of $\pi/R$. To address this issue and, even more importantly, to regularize the singularity in $\Delta^{-2}$ arising from the idealized Bessel vortex state, we take the simplest approach by employing a Gaussian smearing in the transverse momentum $\kappa_1$ of the initial vortex photon:
\begin{align}
  f(\kappa_1;\kappa_0,\sigma_\perp)=N(\kappa_0,\sigma_\perp) \exp\left[{-\frac{(\kappa_1-\kappa_0)^2}{2 \sigma_\perp^2}}\right],
\end{align}
with the normalization factor being given by
\begin{align}
  N^{-2}(\kappa_0,\sigma_\perp)=\frac{1}{2}\sqrt{\pi}\sigma_\perp\left[ 1+ {\rm erf}\left( \frac{\kappa_0}{\sigma_\perp} \right)  \right],
\end{align}
where ${\rm erf}(x)$ is the error function. The previous combination of various factors then becomes the standard $1/(2E_12\omega_1)\cdot[1/(v_1+\cos\vartheta)]$ since the average flux is not affected when the small smearing in the energy is ignored. To summarize, the differential cross section becomes
\begin{align}
  \nonumber
  d\sigma
  &=\delta(q^0)\delta(q^z)\delta_{m_1+m_2+m_e,h_1/2}
  \\
  &\quad\times \frac{\calF}{(2\pi)^32E_12\omega_1(v_1+\cos\vartheta)}
  \frac{dp_2^zd\kappa_e}{2E_2}\frac{dk_2^zd\kappa_2}{2\omega_2},
  \label{eq:dsigma_1}
\end{align}
where
\begin{widetext}
  \begin{align}
    \calF&=\bigg|\int_{\kappa_-}^{\kappa_+} d\kappa_1~f(\kappa_1;\kappa_0,\sigma_\perp)
    \frac{\sqrt{\kappa_1\kappa_2\kappa_e}}{2\Delta}
    \sum_{\eta=\pm}e^{i\eta[m_2\delta_2+(h_1/2-m_e)\delta_e]}
    \calA_{\rm pw}(0,-\eta\delta_2,\eta\delta_e)\bigg|^2,
    \label{eq:Fdef}
  \end{align}
\end{widetext}
and the delimiters are restricted by the triangle condition, $\kappa_\pm=|\kappa_2\pm\kappa_e|$.

We can use the two $\delta$ functions in \cref{eq:dsigma_1} to complete any two of the four final-state variables. The distribution in the final photon energy $\omega_2$ and the cone angle $\theta_2$ is given by
\begin{align}
  \frac{d^2\sigma}{d\omega_2d\theta_2}
  &=
  \frac{\calF}{(2\pi)^3 16E_1\omega_1\kappa_e(v_1+\cos\vartheta)},
  \label{eq:dsigma_2}
\end{align}
where, from now on, the Kronecker delta due to \cref{eq:TAM} is suppressed for brevity.
The correlated constraints in $(\omega_2,\theta_2)$ 
follow from the requirement that the three transverse momenta form a triangle; i.e., $\lambda(\kappa_1^2,\kappa_2^2,\kappa_e^2)\leq 0$, or $(\kappa_1-\kappa_2)^2\leq\kappa_e^2\leq(\kappa_1+\kappa_2)^2$: 
\begin{align}
  \nonumber
  1+\cos(\vartheta-\theta_2)
  &\geq
  \frac{E_1+p_1\cos\vartheta}{\omega_2}
  -\frac{E_1-p_1\cos\theta_2}{\omega_1}  
  \\
  &\geq
  1+\cos(\vartheta+\theta_2).
\end{align}
For a fixed $\theta_2$, the allowed photon energy is $\omega_-\leq\omega_2\leq\omega_+$, where
\begin{align}
  \omega_\pm&
  =\frac{\omega_1(E_1+p_1\cos\vartheta)}
  {\big(E_1-p_1\cos\theta_2\big)+\omega_1\big(1+\cos(\vartheta\pm\theta_2)\big)}.
\end{align}
Alternatively, for a fixed $\omega_2$, the allowed cone angle satisfies
  \begin{align}
    |\gamma(\omega_2)-\beta| \leq \theta_2 \leq \gamma(\omega_2)+\beta.
  \end{align}
  Here $\beta$ is the polar angle of the initial total momentum $\Pvec\equiv\omega_1\hat{\kvec}_1+p_1\hat{\pmb{z}}$ with respect to the $z$ axis,
  \begin{align}
    \beta = \arccos\left[\frac{p_1-\omega_1\cos\vartheta}{P}\right],
  \end{align}
  where $P=\sqrt{(p_1-\omega_1\cos\vartheta)^2+\omega_1^2\sin^2\vartheta}$ is the magnitude of $\Pvec$. Energy-momentum conservation fixes the projection of $\Pvec$ onto $\hat\kvec_2$ to be $C(\omega_2)$, with
  \begin{align}
    C(\omega_2) \equiv E_1+\omega_1-\frac{\omega_1}{\omega_2}(E_1+p_1\cos\vartheta).
  \end{align}
  Thus, $\gamma(\omega_2)=\arccos[C(\omega_2)/P]$ is the angle between the final-photon momentum $\kvec_2$ and the initial total momentum $\Pvec$.

Similarly, the differential cross section in terms of the final-electron energy $E_2$ and cone angle $\theta_e$ is
  \begin{align}
    \frac{d^2\sigma}{dE_2d\theta_e}
    &=
    \frac{\calF}{(2\pi)^3 16E_1\omega_1\kappa_2(v_1+\cos\vartheta)}.
    \label{eq:dsigma_E2}
  \end{align}
  The corresponding correlated kinematic constraint follows from the same
  transverse-triangle condition.  For a fixed $\theta_e$, define
  $C_\pm(\theta_e)\equiv p_1\cos\theta_e-\omega_1\cos(\vartheta\pm\theta_e)$.
  The allowed values of $E_2$, with $p_2=\sqrt{E_2^2-M^2}$, are determined by
  \begin{align}
    C_-p_2
    &\leq(E_1+\omega_1)E_2-M^2
    -\omega_1(E_1+p_1\cos\vartheta)
    \leq C_+p_2.
    \label{eq:electron-energy-double-inequality}
  \end{align}
  When
  $M^2+\omega_1(E_1+p_1\cos\vartheta)\geq M(E_1+\omega_1)$, the allowed range
  is a single interval bounded by the two physical solutions of the equality
  in \cref{eq:electron-energy-double-inequality}. If this condition is
  reversed, the range can, depending on $\theta_e$, split into one or two
  disjoint energy intervals.
  For a fixed $E_2$, the allowed cone angle $\theta_e$ falls in the range
  \begin{align}
    |\gamma_e(E_2)-\beta| \leq \theta_e \leq \gamma_e(E_2)+\beta,
  \end{align}
  where $\gamma_e(E_2)=\arccos[C_e(E_2)/P]$, with
  \begin{align}
    C_e(E_2)\equiv
    \frac{(E_1+\omega_1)E_2-M^2
    -\omega_1(E_1+p_1\cos\vartheta)}{p_2}.
    \label{eq:Ce-definition}
  \end{align}

\section{Numerical results}
\label{sec:numerical_res}

\begin{figure}[htbp]
  \centering
  \includegraphics[width=0.4\textwidth]{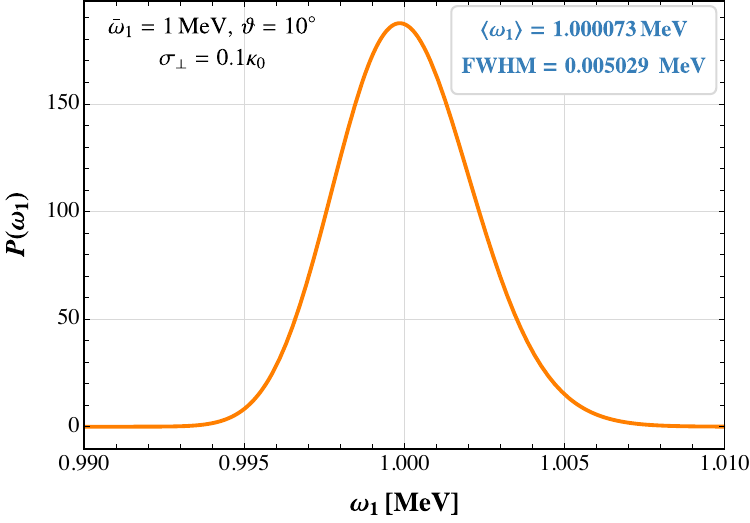}
  \caption{Energy distribution of the initial Bessel-Gaussian vortex photon for the benchmark configuration $(\bar\omega_1,\vartheta)=(1\,\MeV,10^\circ)$ with $\sigma_\perp=0.1\kappa_0$.}
  \label{fig:energy_spread}
\end{figure}

\begin{figure*}[t!]
  \centering
  \includegraphics[width=0.9\textwidth]{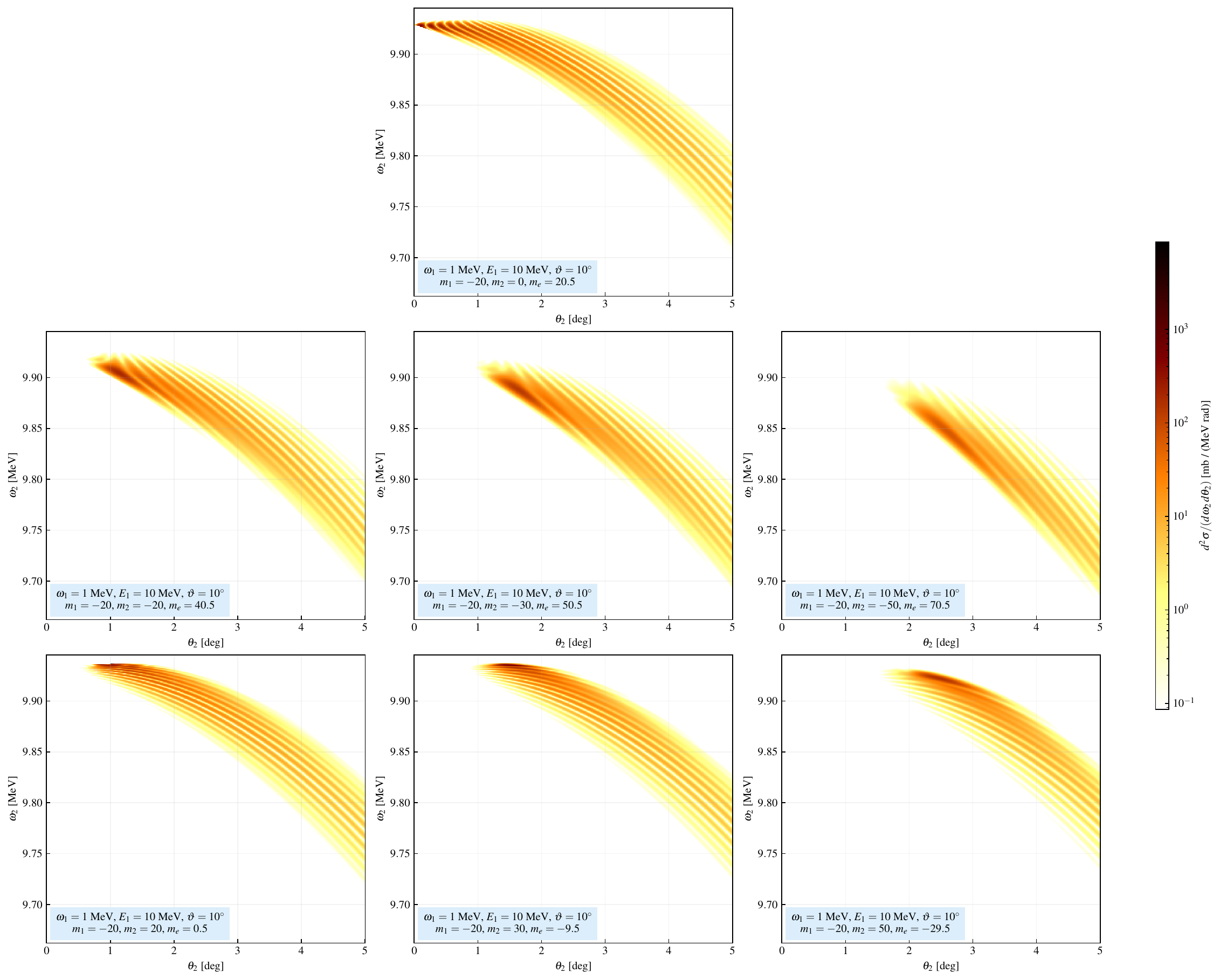}
  \caption{The two-dimensional differential cross section $d^2\sigma/(d\omega_2\,d\theta_2)$ as a function of the final-photon energy $\omega_2$ and cone angle $\theta_2$ for several final-photon topological charges $m_2$. The incident vortex photon has central energy $\bar\omega_1=1\,\MeV$, cone angle $\vartheta=10^\circ$, and topological charge $m_1=-20$, while the incident plane-wave electron has energy $E_1=10\,\MeV$.}
  \label{fig:density_plots}
\end{figure*}

In this section, we proceed with the numerical analysis. We evaluate the differential cross sections in
\cref{eq:dsigma_2,eq:dsigma_E2} for a head-on collision between a Bessel-Gaussian vortex photon propagating along the $-z$ direction and a plane-wave electron propagating along the $+z$ direction. The incident electron energy is fixed at $E_1=10\,\MeV$ and, unless otherwise stated, its helicity is taken to be $h_1/2=+1/2$. For the incident vortex photon, we set $m_1=-20$ and consider two benchmark configurations, $(\bar\omega_1,\vartheta) = (1\,\MeV,10^\circ)$ and $(10\,\keV,1^\circ)$. The former illustrates the production of multi-MeV vortex photons from a MeV-scale incident photon, while the latter represents an X-ray vortex source with a smaller and experimentally more accessible opening angle. For a specified final-photon topological charge $m_2$, the charge of the final vortex electron is fixed by the conservation law of the total angular momentum on the $z$ axis, i.e. \cref{eq:TAM}.
To regularize the singular behavior of an ideal Bessel state, we use the Bessel-Gaussian profile introduced in the previous section and choose $\sigma_\perp=0.1\kappa_0$ for both benchmarks. As the integral of $[f(\kappa_1;\kappa_0,\sigma_\perp)]^2$ over $\kappa_1$ is normalized to unity, the distribution in energy $\omega_1$ is, $(\omega_1/\kappa_1)[f(\kappa_1;\kappa_0,\sigma_\perp)]^2$ with $\kappa_1=\omega_1\sin\vartheta$ and $\kappa_0=\bar\omega_1\sin\vartheta$, where $\bar\omega_1$ is the central value of energy. For the $(1\,\MeV,10^\circ)$ benchmark, \cref{fig:energy_spread} shows that monochromaticity is guaranteed at the per mil level. The energy spread is even smaller for the $(10\,\keV,1^\circ)$ benchmark. It is therefore sufficient to evaluate the flux and external kinematic factors at the central value $\bar\omega_1$ of energy, while retaining the full $\kappa_1$ dependence in the Bessel-Gaussian convolution.

The following one-dimensional spectra are obtained by integrating the two-dimensional distributions over their kinematically allowed
regions:
\begin{align}
  \frac{d\sigma}{d\theta_2}
  &=\int_{\omega_-(\theta_2)}^{\omega_+(\theta_2)}
  d\omega_2\,
  \frac{d^2\sigma}{d\omega_2d\theta_2},
  \\
  \frac{d\sigma}{d\omega_2}
  &=\int_{\theta_{2,\min}(\omega_2)}^{\theta_{2,\max}(\omega_2)}
  d\theta_2\,
  \frac{d^2\sigma}{d\omega_2d\theta_2},
\end{align}
with analogous integrations over $E_2$ and $\theta_e$ for the final
vortex electron, while we have considered the smearing effects for the boundaries in the numerical computation.  
Unless otherwise stated, we fix the initial-electron helicity at
$h_1/2=+1/2$, but average or sum over the helicities of all other particles:
\begin{align}
  d\sigma_{\rm unpol.}
  =\frac{1}{2}
  \sum_{\lambda_1,\lambda_2,h_2}
  d\sigma(\lambda_1 h_1;\lambda_2 h_2).
  \label{eq:unpol}
\end{align}
The only exception is the helicity decomposition shown in
\cref{fig:helicity}, where the curves labeled `All' denote the direct
helicity sum without the initial-photon averaging factor.

\subsection{Distributions for final vortex photon}

%
\begin{figure*}[t!]
  \centering
  \includegraphics[width=0.465\textwidth]{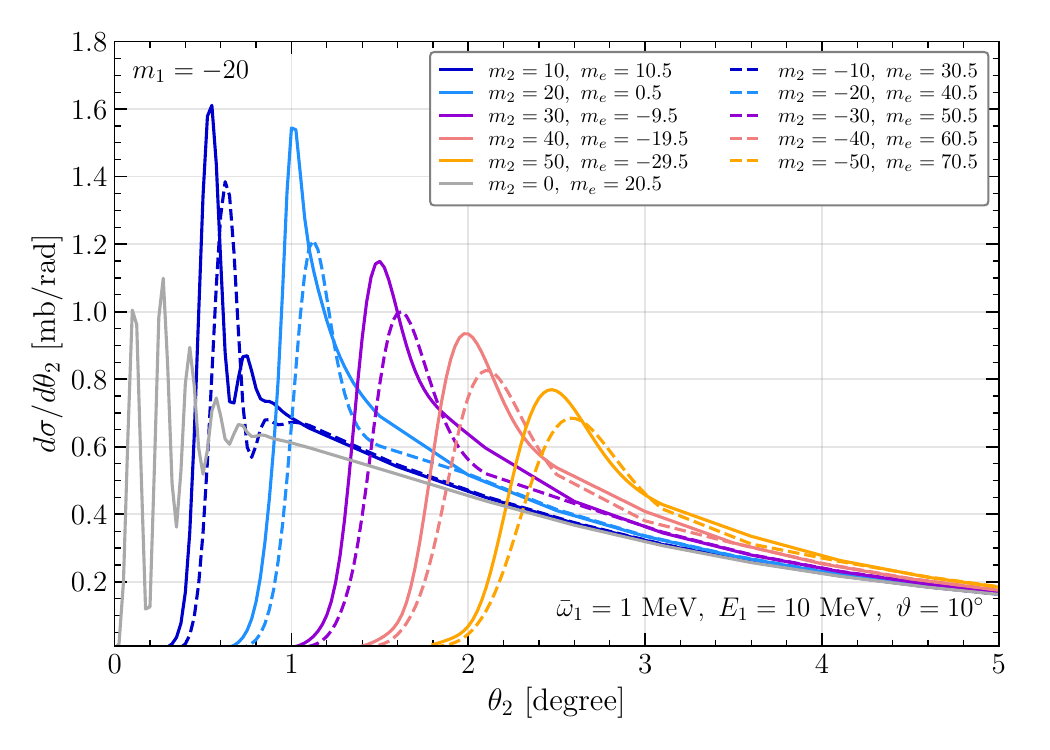}
  \includegraphics[width=0.49\textwidth]{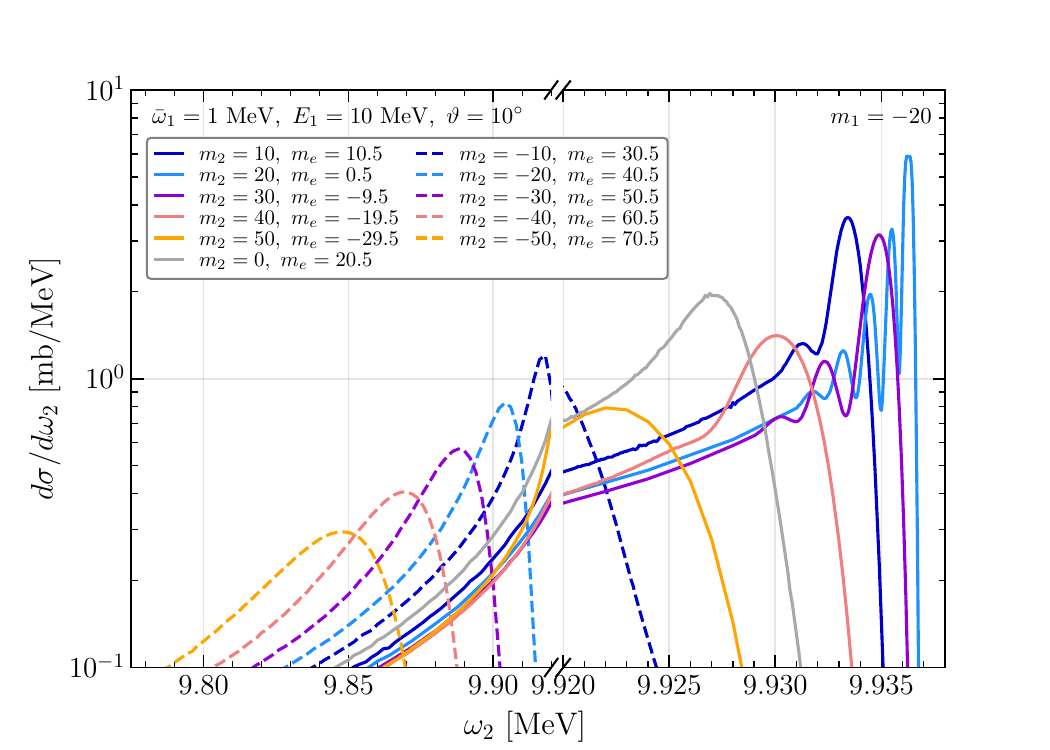}
  \caption{The one-dimensional differential cross section $d\sigma/d\theta_2$ ($d\sigma/d\omega_2$) as a function of $\theta_2$ ($\omega_2$) in the left (right) panel. The parameters for the incident photon and electron are the same as those in \cref{fig:density_plots}.}
  \label{fig:spectrum_1D_omega2_theta2_1MeV}
\end{figure*}

\begin{figure}[t!]
  \centering
  \includegraphics[width=0.48\textwidth]{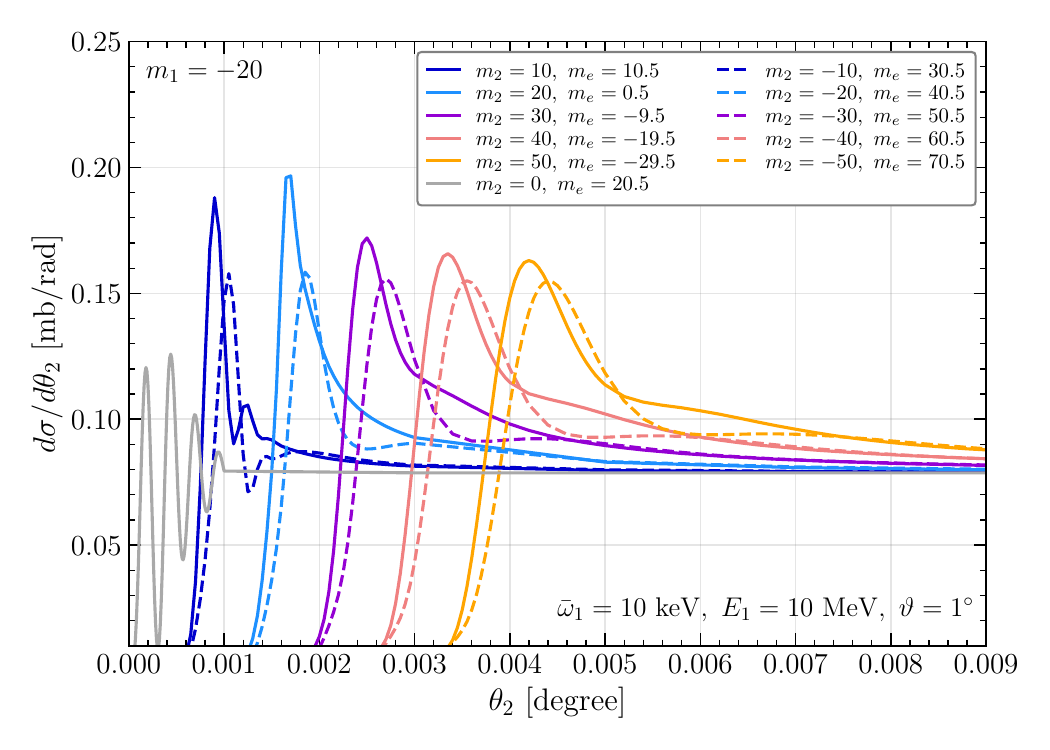}
  \caption{The one-dimensional differential cross section $d\sigma/d\theta_2$ as a function of $\theta_2$ for $\bar\omega_1=10~\keV$ and $\vartheta=1^\circ$. The other parameters for the incident photon and electron are the same as those in \cref{fig:density_plots}.}
  \label{fig:spectrum_1D_theta2_10keV}
\end{figure}
\begin{figure}[t!]
  \centering
  \includegraphics[width=0.48\textwidth]{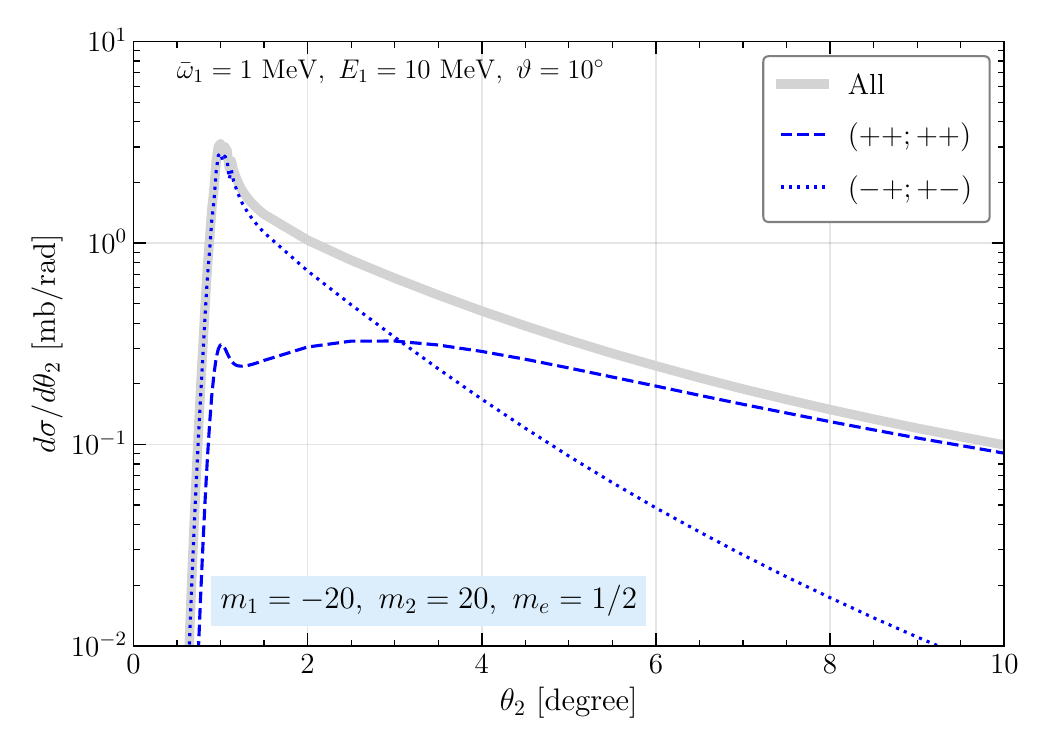}
  \par\vspace{-0.5em}
  \includegraphics[width=0.48\textwidth]{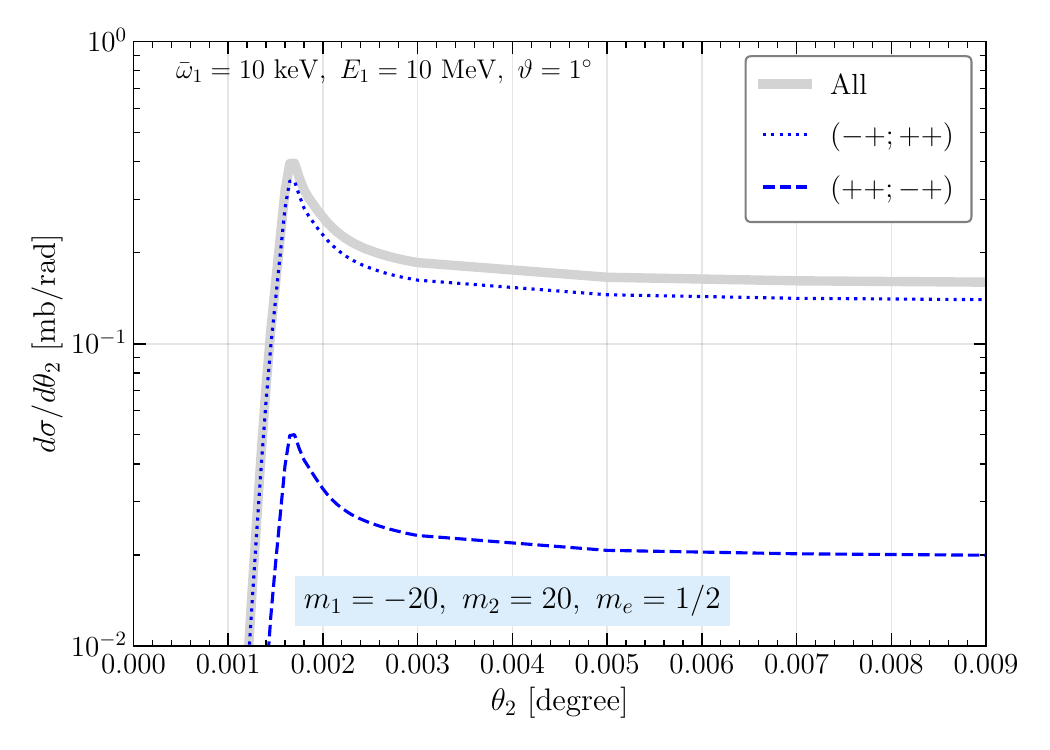}
  \caption{The contributions to $d\sigma/d\theta_2$ from individual helicity channels for $\bar\omega_1=1$ MeV (upper panel) and $\bar\omega_1=10$ keV (lower).}
  \label{fig:helicity}
\end{figure}

We first consider the MeV-scale benchmark $(\bar\omega_1,\vartheta)=(1~\MeV,10^\circ)$. This energy regime is motivated by the recent experimental indication of vortex-photon generation at sub-MeV energies \cite{Wei:2025zsv}. Figure~\ref{fig:density_plots} shows the unpolarized differential cross section $d\sigma/(d\omega_2d\theta_2)$ defined in \cref{eq:unpol} for $m_2=0,\,\pm20,\,\pm30$, and $\pm50$, and the corresponding $m_e$ is fixed by \cref{eq:TAM}. We restrict the presentation to $\theta_2\leq5^\circ$, because the dependence on $m_2$ becomes weaker at larger angles. The allowed events populate narrow, curved bands in the $(\theta_2,\omega_2)$ plane, demonstrating the strong correlation between the final-photon opening angle and energy: fixing either variable confines the other to a narrow interval. Within each band, the oscillatory fringes arise from interference between the two transverse-momentum configurations labeled by $\eta=\pm$. Their patterns depend on the relative signs of the initial and final topological charges, with $m_1m_2>0$ and $m_1m_2<0$ displayed in the second and third rows of \cref{fig:density_plots}, respectively.
Each distribution contains a localized region in which the cross section is the largest, and its position varies systematically with $m_2$. For a pair of opposite charges $\pm m_2$, the dominant regions occur at similar values of $\theta_2$, whereas the spectrum for positive $m_2$ is slightly shifted towards higher $\omega_2$. Increasing $|m_2|$ moves the dominant region to larger opening angles and, through the energy-angle correlation, to lower photon energies.

The one-dimensional projections defined above are shown in \cref{fig:spectrum_1D_omega2_theta2_1MeV}. The angular spectra are suppressed near $\theta_2=0$, rise rapidly to an $m_2$-dependent maximum, and then decrease toward a nearly common large-angle behavior. The oscillations are most pronounced for $m_2=0$, while the principal peaks move to larger $\theta_2$ as $|m_2|$ increases, consistent with the two-dimensional distributions. To describe the energy spectra, it is useful to define $\Delta m\equiv m_2-(-m_1)=m_1+m_2$. The sign in this definition accounts for the fact that the incident photon propagates along the $-z$ direction. The hardest spectrum occurs for $\Delta m=0$, namely $m_2=-m_1=20$. As $|\Delta m|$ increases, the spectrum becomes progressively softer, with a mild asymmetry between positive and negative $\Delta m$.
Although a direct measurement of the helical wavefront of a high-energy photon remains challenging, the charge-dependent separation in $\theta_2$ and $\omega_2$ offers an indirect handle on its vortex state. Angular or energy selection could therefore be used to identify kinematic regions enriched in particular values of $m_2$.

We next consider the X-ray benchmark $(\bar\omega_1,\vartheta)=(10~\keV,1^\circ)$, for which the incident vortex photon is more readily accessible experimentally. As shown in \cref{fig:spectrum_1D_theta2_10keV}, the same charge-dependent peak structure persists but is compressed into the much smaller angular range $\theta_2\lesssim0.009^\circ$. Within this region, fixing $\theta_2$ also restricts the final-photon energy to a narrow kinematic interval. Thus, the correlation between $m_2$, $\theta_2$, and $\omega_2$ is not specific to the MeV-scale benchmark, although resolving it in the X-ray case requires substantially finer angular and energy resolution.

Finally, we examine the helicity composition of the produced vortex photon. In \cref{fig:helicity}, we take $m_1=-20$, $m_2=20$, and $m_e=1/2$ for a positively polarized incident electron, $h_1=+1$. A channel label $(\lambda_1h_1;\lambda_2h_2)$ specifies the helicities of the initial and final photon--electron pairs. As noted above, the curves labeled `All' are direct sums over the helicity channels and do not contain the factor $1/2$ associated with averaging over the initial-photon helicity. For $\bar\omega_1=1~\MeV$, the dominant contributions are $(++;++)$ and $(-+;+-)$. The fully helicity-conserving channel $(++;++)$ dominates for $\theta_2\gtrsim4^\circ$, whereas the electron-helicity-flip channel $(-+;+-)$ dominates for $\theta_2\lesssim2^\circ$, where the final electron is backscattered. For $\bar\omega_1=10~\keV$, the larger electron-photon energy hierarchy keeps the final electron in the forward direction, and the $(-+;++)$ channel dominates throughout the display range. In all of these leading channels, the final photon has a positive helicity, matching the helicity label $h_1=+1$ of the incident polarized electron.

In summary, triple-vortex backward Compton scattering simultaneously increases the photon energy and produces final states whose angular and energy spectra depend on their topological charges. The separation of the $m_2$-dependent peaks provides a basis for kinematic post-selection of vortex-photon states, while the helicity decomposition shows a strong preference for final photons whose helicity follows that of a longitudinally polarized electron beam. These features persist for both the MeV-scale and X-ray benchmark configurations.

\subsection{Distributions for final vortex electron}
%
\begin{figure*}[t!]
  \centering
  \includegraphics[width=0.99\textwidth]{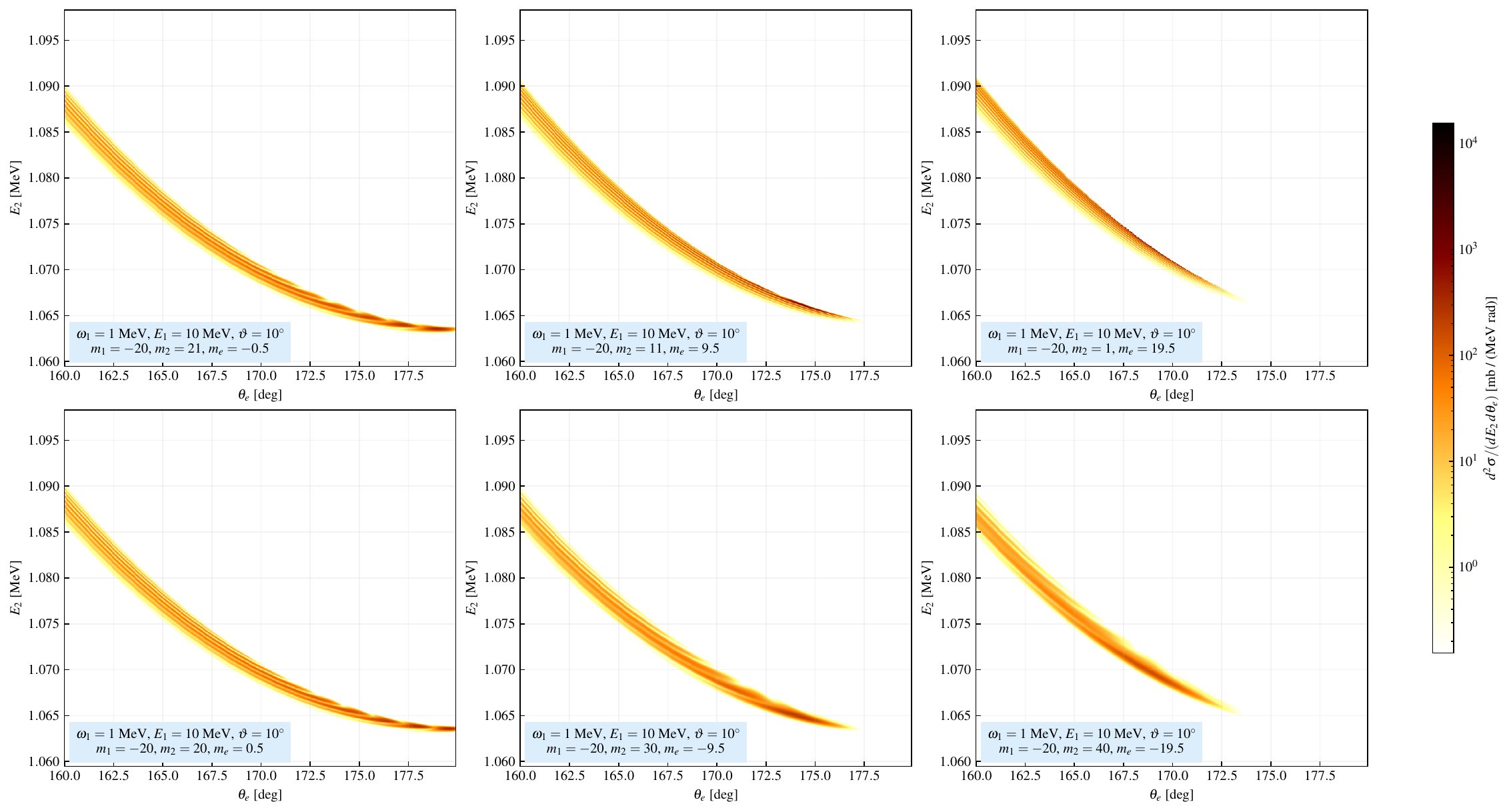}

  \caption{The two-dimensional differential cross section $d^2\sigma/(dE_2d\theta_e)$ as a function of $E_2$ and $\theta_e$ for different topological charges.}
  \label{fig:density_plots-electron}
\end{figure*}
\begin{figure*}[t!]
  \centering
  \includegraphics[width=0.48\textwidth]{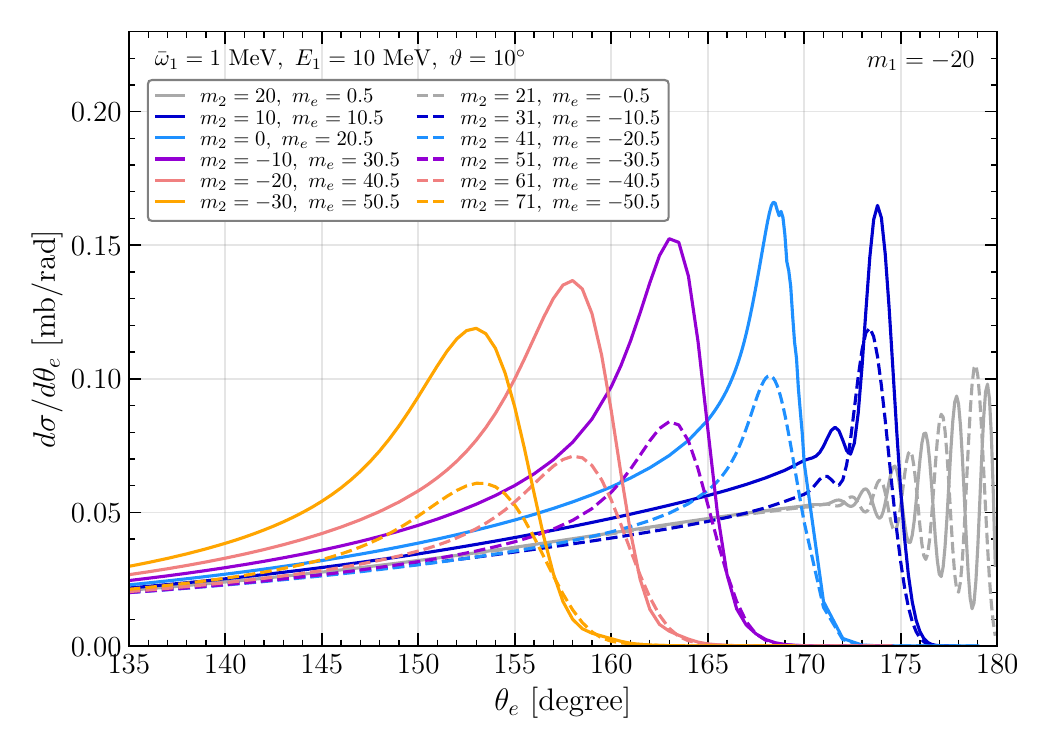}
  \includegraphics[width=0.48\textwidth]{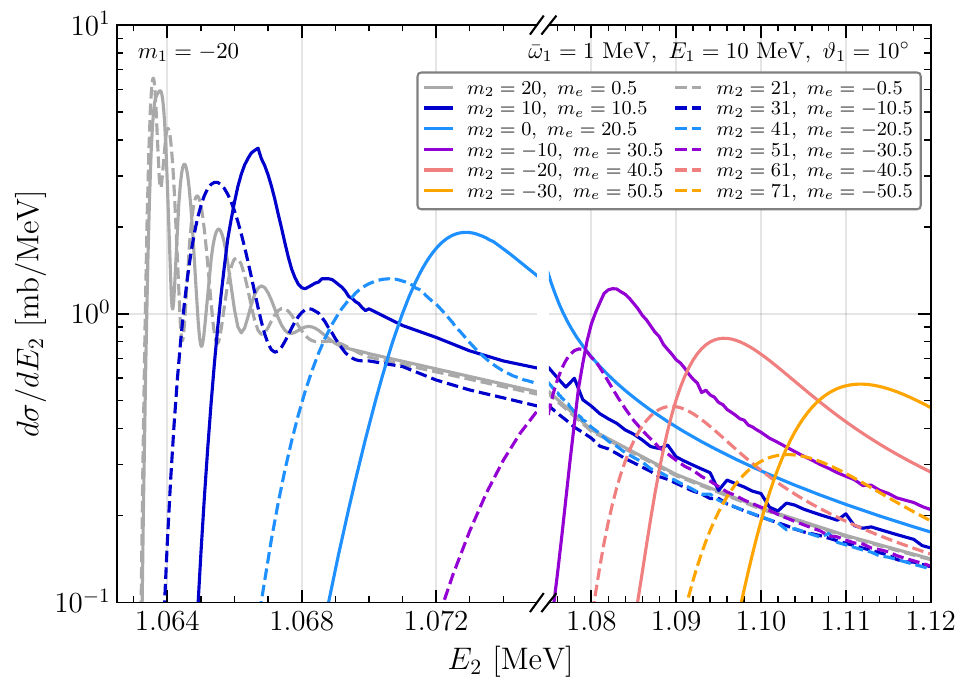}
  \caption{The one-dimensional differential cross section $d\sigma/d\theta_e$ ($d\sigma/dE_2$) as a function of $\theta_e$ ($E_2$) in the left (right) panel.}
  \label{fig:spectrum_1D_E2_thetae_10deg}
\end{figure*}

We now turn to the final-state vortex electron for the MeV-scale benchmark $(\bar\omega_1,\vartheta)=(1~\MeV,10^\circ)$. Figure~\ref{fig:density_plots-electron} shows the unpolarized differential cross section $d\sigma/(dE_2d\theta_e)$ defined in \cref{eq:unpol} for representative values of $(m_2,m_e)$. The final electron is emitted predominantly toward the $-z$ direction, so its cone angle about that direction is written more transparently as $\alpha_e\equiv\pi-\theta_e$. Thus, the displayed interval $160^\circ\leq\theta_e\leq180^\circ$ corresponds to $0^\circ\leq\alpha_e\leq20^\circ$. As in photon distributions, the allowed events populate narrow, curved bands in the $(\theta_e,E_2)$ plane and exhibit interference fringes generated by the $\eta=\pm$ components. The detailed fringe pattern depends on both the magnitude and sign of $m_e$.

For the smallest charges, $m_e=\pm1/2$, several localized maxima are visible along the kinematic band. After integration over one of the two variables, these maxima produce the damped oscillations in $d\sigma/d\theta_e$ and $d\sigma/dE_2$ shown in \cref{fig:spectrum_1D_E2_thetae_10deg}. As $|m_e|$ increases, the dominant angular peak moves toward a smaller $\theta_e$, or equivalently toward a larger cone angle $\alpha_e$, while the corresponding energy distribution shifts slightly toward larger $E_2$. The one-dimensional projections therefore retain the same systematic charge dependence seen in the density plots. For the benchmark considered here, the dominant events contain electrons with total energies around $1.06$--$1.09~\MeV$, cone angles that reach several tens of degrees, and topological charges of the order $|m_e|\sim50$. Backward Compton scattering can therefore produce MeV-scale vortex electrons with simultaneously sizable opening angles and large angular-momentum projections, while angular and energy selection provides kinematic handles for distinguishing different $m_e$ sectors.

\begin{figure}[t!]
  \centering
  \includegraphics[width=0.48\textwidth]{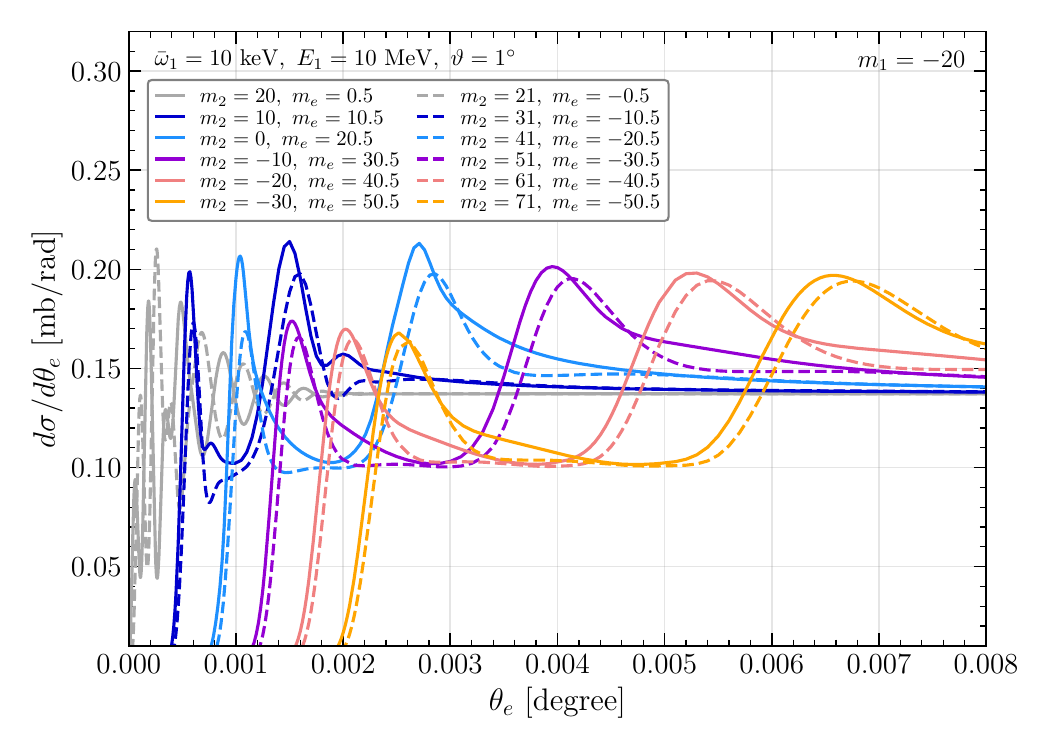}
  \caption{The one-dimensional differential cross section $d\sigma/d\theta_e$ as a function of $\theta_e$ for $\bar\omega_1=10~\keV$ and $\vartheta=1^\circ$.}
  \label{fig:spectrum_1D_thetae_10keV}
\end{figure}

The differential cross section $d\sigma/d\theta_e$ for $(\bar\omega_1,\vartheta)=(10~\keV,1^\circ)$ is shown in \cref{fig:spectrum_1D_thetae_10keV}. As discussed previously, in this configuration the final vortex electron propagates predominantly in the forward direction. Consequently, its opening angle $\theta_e$ is small despite its relatively large energy. In addition, the distributions exhibit a characteristic two-peak structure, which originates from the two disjoint kinematically allowed energy intervals of the final electron. This two-peak structure complicates the selection of a definite topological charge.

The correlated charge-dependent features of the photon and electron spectra admit a unified semiclassical interpretation in terms of a common transverse-radius condition.  A Bessel mode of orbital order $\ell$ and transverse momentum $\kappa$ has a characteristic caustic radius $\rho_\perp\sim|\ell|/\kappa$, so transverse overlap is enhanced when $|m_1|/\kappa_1\sim|m_2|/\kappa_2\sim|\ell_e|/\kappa_e$, where $\ell_e\simeq m_e$ for large electron charges, up to spin-dependent shifts. This condition explains why increasing $|m_2|$ moves the photon peak toward larger $\theta_2$ and why increasing $|m_e|$ similarly favors a larger electron cone angle. Together with the angular-momentum selection rule and energy-momentum conservation, it also accounts for the correlated softening of the photon and hardening of the electron spectra, while the coherent $\eta=\pm$ contributions superimpose the observed interference fringes on this overall envelope.

\section{Conclusion}
\label{sec:conclusion}

We have developed a theoretical description of backward Compton scattering with three vortex particles, $\gamma_\vx+e^-_\pw\to\gamma_\vx+e^-_\vx$. The initial and final photons and the final electron are described by vortex states, while the incident electron is a plane wave. The azimuthal integrations reduce the transverse-momentum conservation condition to two geometric configurations whose coherent sum generates the interference structure characteristic of the process. Rotational symmetry about the collision axis leads to the exact selection rule $h_1/2-m_1=m_2+m_e$, which directly relates the angular-momentum projections of the two final vortex particles to those of the incident states. The incident vortex photon is further described by a Bessel-Gaussian wave packet, thereby obtaining a physically normalized cross section and regularizing the singular behavior at the boundaries of the transverse-momentum triangle.

Our numerical studies are performed for a $10~\MeV$ electron colliding with a vortex photon of central energy $1~\MeV$ or $10~\keV$, including the distributions of the final vortex photon and electron, as well as the dependence of their opening angles and energies on the corresponding topological charges.

The final-state vortex photon exhibits correlated energy and cone angle distributions into a curved band. The coherent $\eta=\pm$ contributions generate charge-dependent interference fringes, while the dominant regions shift systematically with the final-photon charge. Increasing $|m_2|$ moves the angular peak to a larger cone angle, and the hardest photon spectrum occurs near a vanishing angular-momentum transfer, $m_2=-m_1$. Although most results are shown for an MeV incident vortex photon, the same qualitative behavior persists for the $10~\keV$ incident photon benchmark. The helicity decomposition further shows that a longitudinally polarized incident electron produces a strong preference for a final-photon helicity correlated with the electron polarization. These energy, cone angle, and helicity correlations provide practical handles for enhancing the capability of separating vortex topological sectors through kinematic post-selection.

The $1~\MeV$ incident photon benchmark also accesses a deep-recoil region in which the final electron is emitted predominantly along the $-z$ direction with a total energy near $1~\MeV$.  Its dominant peak moves toward a larger opening angle and slightly higher energy as $|m_e|$ increases, allowing vortex-electron states with cone angles of several tens of degrees and angular-momentum projections of order tens.  The charge-dependent peak locations of both final particles are consistent with a common transverse-radius argument.

Our results therefore identify triple-vortex backward Compton scattering as a possible mechanism for transferring energy, angular momentum, and polarization into correlated high-energy vortex states.  Establishing its experimental reach will require extending the present treatment to more realistic circumstances, including beam alignment, detector resolution, as well as developing observables that can resolve or efficiently post-select the final OAM sectors, to some of which we will come back in the future work.

\acknowledgments
This work was supported in part by the Grants
No.\,NSFC-12605178 and
No.\,NSFC-12035008.

\onecolumngrid
\appendix

\section{Helicity amplitudes and differential cross sections for general plane-wave Compton scattering}
\label{sec:helicity_amp}
The bispinor wavefunctions for the initial electron moving in the $+z$ axis and the final electron in any direction are
\begin{subequations}
  \begin{align}
    & u(p_1,h_1)=\left(
      \begin{array}{r}
        \epsilon_{1+} w(p_1,h_1) \\
        h_1 \epsilon_{1-} w(p_1,h_1)
    \end{array}\right) ,\quad w(p_1,h_1)= \frac{1}{\sqrt{2}}\left(
      \begin{array}{r}
        h_1\sqrt{1+h_1}\\ \sqrt{1-h_1}
    \end{array}\right),
    \\
    & u(p_2,h_2)=\left(
      \begin{array}{r}
        \epsilon_{2+} w(p_2,h_2) \\
        h_2 \epsilon_{2-} w(p_2,h_2)
    \end{array}\right) ,\quad w(p_2,h_2)= \frac{1}{\sqrt{2}}\left(
      \begin{array}{r}
        h_2 e^{-i\varphi_e/2}\sqrt{1+h_2 \cos\theta_e} \\
        e^{+i\varphi_e/2}\sqrt{1-h_2 \cos\theta_e}
    \end{array}\right),
  \end{align}
\end{subequations}
where $\epsilon_{a\pm}=\sqrt{E_a\pm M}$, $v_1=|\pvec_1|/E_1$, and $\tau_a=|\pvec_1|/\omega_a$ with $a=1~(2)$ denoting the initial (final) electron and photon. The polarization vectors for the initial and final photons in any direction are
\begin{subequations}
  \begin{align}
    &\vec\epsilon_a
    =\vec\eta_{(\lambda_a)}e^{i\bar\lambda_a\varphi_a}\cos^2\frac{\theta_a}{2}
    +\vec\eta_{(\bar\lambda_a)}e^{+i\lambda_a\varphi_a}\sin^2\frac{\theta_a}{2}
    +\vec\eta_{(0)}\frac{\lambda_a}{\sqrt{2}}\sin\theta_a,
    \\
    &\vec\eta_{(\lambda)}=-\frac{1}{\sqrt{2}}(\lambda,i,0)^T,~
    \vec\eta_{(0)}=(0,0,1)^T.
  \end{align}
\end{subequations}

The amplitude for general plane-wave scattering is given by
\begin{align}
  \calA_\pw&=\frac{e^2}{|\pvec_1|}\left(
    -\frac{A}{v_1^{-1}-\hat{\pvec}_1\cdot\hat{\kvec}_1}
  +\frac{B}{v_1^{-1}-\hat{\pvec}_1\cdot\hat{\kvec}_2}\right).
\end{align}
where
\begin{subequations}
  \begin{align}
    \nonumber
    2\omega_1A &= \bar{u}(p_2,h_2) \slashed{\epsilon}_2^* (\slashed{p}_1+\slashed{k}_1+M)\slashed{\epsilon}_1 u(p_1,h_1),
    \\
    2\omega_2B &= \bar{u}(p_2,h_2) \slashed{\epsilon}_1 (\slashed{p}_1-\slashed{k}_2+M) \slashed{\epsilon}_2^* u(p_1,h_1).
  \end{align}
\end{subequations}
We introduce the following short-cuts:
\begin{subequations}
  \begin{align}
    &c_a=\cos\frac{\theta_a}{2},\quad s_a=\sin\frac{\theta_a}{2},\quad c_e=\cos\frac{\theta_e}{2}, \quad s_e=\sin\frac{\theta_e}{2},
    \\
    &z_a=e^{i\varphi_a},\quad z_e=e^{i\varphi_e/2},
    \\
    &\xi_{\pm}= \epsilon_{1-}\epsilon_{2+}\pm\epsilon_{1+}\epsilon_{2-},\quad
    \xi_{\eta_1 \eta_2}= (\epsilon_{1-}+\eta_1 \epsilon_{1+})(\epsilon_{2-}+\eta_2 \epsilon_{2+}).
  \end{align}
\end{subequations}
Observing that the helicity amplitudes satisfy the relations
\begin{subequations}
  \begin{align}
    &A^*(\lambda_1h_1;\lambda_2h_2)=h_1h_2A(\bar\lambda_1\bar h_1;\bar \lambda_2\bar h_2),
    \\
    &B^*(\lambda_1h_1;\lambda_2h_2)=h_1h_2B(\bar\lambda_1\bar h_1;\bar \lambda_2\bar h_2),
  \end{align}
\end{subequations}
with $\bar\lambda_a=-\lambda_a$ and $\bar h_a=-h_a$, only one half of helicity amplitudes are explicitly presented below. The eight $A$ terms are
\begin{subequations}
  \begin{align}
    \nonumber
    &A(++;++) = 2\tau_1 \Big(
      c_1 c_2 c_e s_1 s_2 z_e
      - c_1 c_2^2 s_1 s_e z_2z_e^*
    \Big)\xi_+
    \\
    & \qquad+\Big(
      c_1 c_2 c_e s_1 s_2 z_e
      + c_e s_1^2 s_2^2 z_1z_2^*z_e
      - c_1 c_2^2 s_1 s_e z_2z_e^*
      - c_2 s_1^2 s_2 s_e z_1z_e^*
    \Big)\xi_{++},
    \\
    \nonumber
    &A(+-;++) = 2\tau_1 \Big(
      c_1 c_2 s_1 s_2 s_e z_e^*
      - c_1 c_e s_1 s_2^2 z_2^*z_e
    \Big)\xi_-
    \\
    & \qquad+\Big(
      c_1 c_e s_1 s_2^2 z_2^*z_e
      + c_1^2 c_2 c_e s_2 z_1^*z_e
      - c_1 c_2 s_1 s_2 s_e z_e^*
      - c_1^2 c_2^2 s_e z_1^*z_2z_e^*
    \Big)\xi_{-+},
    \\
    \nonumber
    &A(++;+-) = 2\tau_1 \Big(
      - c_1 c_2 s_1 s_2 s_e z_e
      - c_1 c_2^2 c_e s_1 z_2z_e^*
    \Big)\xi_-
    \\
    & \qquad+\Big(
      c_1 c_2 s_1 s_2 s_e z_e
      + c_1 c_2^2 c_e s_1 z_2z_e^*
      + c_2 c_e s_1^2 s_2 z_1z_e^*
      + s_1^2 s_2^2 s_e z_1z_2^*z_e
    \Big)\xi_{+-},
    \\
    \nonumber
    &A(-+;++) = 2\tau_1 \Big(
      c_1 c_2^2 s_1 s_e z_2 z_e^*
      - c_1 c_2 c_e s_1 s_2 z_e
    \Big)\xi_+
    \\
    & \qquad+\Big(
      c_1 c_2^2 s_1 s_e z_2z_e^*
      + c_1^2 c_e s_2^2 z_1z_2^*z_e
      - c_1 c_2 c_e s_1 s_2 z_e
      - c_1^2 c_2 s_2 s_e z_1z_e^*
    \Big)\xi_{--},
    \\
    \nonumber
    &A(++;-+) = 2\tau_1 \Big(
      - c_1 c_2 c_e s_1 s_2 z_e
      - c_1 s_1 s_2^2 s_e z_2z_e^*
    \Big)\xi_+
    \\
    & \qquad+\Big(
      c_2 s_1^2 s_2 s_e z_1z_e^*
      + c_2^2 c_e s_1^2 z_1z_2^*z_e
      - c_1 c_2 c_e s_1 s_2 z_e
      - c_1 s_1 s_2^2 s_e z_2z_e^*
    \Big)\xi_{++},
    \\
    \nonumber
    &A(+-;+-) = 2\tau_1 \Big(
      c_1 c_2 c_e s_1 s_2 z_e^*
      + c_1 s_1 s_2^2 s_e z_2^*z_e
    \Big)\xi_+
    \\
    & \qquad+\Big(
      c_1 c_2 c_e s_1 s_2 z_e^*
      + c_1 s_1 s_2^2 s_e z_2^*z_e
      + c_1^2 c_2 s_2 s_e z_1^*z_e
      + c_1^2 c_2^2 c_e z_1^*z_2z_e^*
    \Big)\xi_{--},
    \\
    \nonumber
    &A(++;--) = 2\tau_1 \Big(
      c_1 c_2 s_1 s_2 s_e z_e
      - c_1 c_e s_1 s_2^2 z_2z_e^*
    \Big)\xi_-
    \\
    & \qquad+\Big(
      c_1 c_e s_1 s_2^2 z_2z_e^*
      + c_2^2 s_1^2 s_e z_1z_2^*z_e
      - c_1 c_2 s_1 s_2 s_e z_e
      - c_2 c_e s_1^2 s_2 z_1z_e^*
    \Big)\xi_{+-},
    \\
    \nonumber
    &A(+-;-+) = 2\tau_1 \Big(
      - c_1 c_2 s_1 s_2 s_e z_e^*
      - c_1 c_2^2 c_e s_1 z_2^*z_e
    \Big)\xi_-
    \\
    & \qquad+\Big(
      c_1 c_2 s_1 s_2 s_e z_e^*
      + c_1 c_2^2 c_e s_1 z_2^*z_e
      - c_1^2 c_2 c_e s_2 z_1^*z_e
      - c_1^2 s_2^2 s_e z_1^*z_2z_e^*
    \Big)\xi_{-+}.
  \end{align}
\end{subequations}
The eight $B$ terms are
\begin{subequations}
  \begin{align}
    \nonumber
    &B(++;++) = 2\tau_2 \Big(
      c_1 c_2 c_e s_1 s_2 z_e
      + c_2 s_1^2 s_2 s_e z_1z_e^*
    \Big)\xi_+
    \\
    & \qquad+\Big(
      - c_1 c_2 c_e s_1 s_2 z_e
      - c_1 c_2^2 s_1 s_e z_2z_e^*
      - c_2 s_1^2 s_2 s_e z_1z_e^*
      - c_1^2 c_2^2 c_e z_1^*z_2z_e
    \Big)\xi_{--},
    \\
    \nonumber
    &B(+-;++) = 2\tau_2 \Big(
      c_1 c_2 s_1 s_2 s_e z_e^*
      + c_1^2 c_2 c_e s_2 z_1^*z_e
    \Big)\xi_-
    \\
    & \qquad+\Big(
      c_1 c_2 s_1 s_2 s_e z_e^*
      + c_1 c_e s_1 s_2^2 z_2^*z_e
      + c_1^2 c_2 c_e s_2 z_1^*z_e
      + s_1^2 s_2^2 s_e z_1z_2^*z_e^*
    \Big)\xi_{+-},
    \\
    \nonumber
    &B(++;+-) = 2\tau_2 \Big(
      c_2 c_e s_1^2 s_2 z_1z_e^*
      - c_1 c_2 s_1 s_2 s_e z_e
    \Big)\xi_-
    \\
    & \qquad+\Big(
      c_1 c_2^2 c_e s_1 z_2z_e^*
      + c_2 c_e s_1^2 s_2 z_1z_e^*
      - c_1 c_2 s_1 s_2 s_e z_e
      - c_1^2 c_2^2 s_e z_1^*z_2z_e
    \Big)\xi_{-+},
    \\
    \nonumber
    &B(-+;++) = 2\tau_2 \Big(
      c_1^2 c_2 s_2 s_e z_1z_e^*
      - c_1 c_2 c_e s_1 s_2 z_e
    \Big)\xi_+
    \\
    & \qquad+\Big(
      c_1 c_2 c_e s_1 s_2 z_e
      + c_1 c_2^2 s_1 s_e z_2z_e^*
      - c_1^2 c_2 s_2 s_e z_1z_e^*
      - c_2^2 c_e s_1^2 z_1^*z_2z_e
    \Big)\xi_{--},
    \\
    \nonumber
    &B(++;-+) = 2\tau_2 \Big(
      - c_1 c_2 c_e s_1 s_2 z_e
      - c_2 s_1^2 s_2 s_e z_1z_e^*
    \Big)\xi_+
    \\
    & \qquad+\Big(
      c_1 c_2 c_e s_1 s_2 z_e
      + c_2 s_1^2 s_2 s_e z_1z_e^*
      - c_1 s_1 s_2^2 s_e z_2z_e^*
      - c_1^2 c_e s_2^2 z_1^*z_2z_e
    \Big)\xi_{++},
    \\
    \nonumber
    &B(+-;+-) = 2\tau_2 \Big(
      c_1 c_2 c_e s_1 s_2 z_e^*
      - c_1^2 c_2 s_2 s_e z_1^*z_e
    \Big)\xi_+
    \\
    & \qquad+\Big(
      c_1 s_1 s_2^2 s_e z_2^*z_e
      + c_1^2 c_2 s_2 s_e z_1^*z_e
      - c_1 c_2 c_e s_1 s_2 z_e^*
      - c_e s_1^2 s_2^2 z_1z_2^*z_e^*
    \Big)\xi_{++},
    \\
    \nonumber
    &B(++;--) = 2\tau_2 \Big(
      c_1 c_2 s_1 s_2 s_e z_e
      - c_2 c_e s_1^2 s_2 z_1z_e^*
    \Big)\xi_-
    \\
    & \qquad+\Big(
      c_1 c_2 s_1 s_2 s_e z_e
      + c_1 c_e s_1 s_2^2 z_2z_e^*
      - c_2 c_e s_1^2 s_2 z_1z_e^*
      - c_1^2 s_2^2 s_e z_1^*z_2z_e
    \Big)\xi_{+-},
    \\
    \nonumber
    &B(+-;-+) = 2\tau_2 \Big(
      - c_1 c_2 s_1 s_2 s_e z_e^*
      - c_1^2 c_2 c_e s_2 z_1^*z_e
    \Big)\xi_-
    \\
    & \qquad+\Big(
      c_1 c_2^2 c_e s_1 z_2^*z_e
      + c_2^2 s_1^2 s_e z_1z_2^*z_e^*
      - c_1 c_2 s_1 s_2 s_e z_e^*
      - c_1^2 c_2 c_e s_2 z_1^*z_e
    \Big)\xi_{-+}.
  \end{align}
\end{subequations}
In the ultrarelativistic limit one half of these $A$ and $B$ terms approaches zero.

The differential cross section is
\begin{align}
  d\sigma=\frac{1}{4p_1\cdot k_1}
  |\calA_\pw|^2(2\pi)^4\delta^4(p_1+k_1-p_2-k_2)
  \frac{d^3\pvec_2}{(2\pi)^32E_2}\frac{d^3\kvec_2}{(2\pi)^32\omega_2},
\end{align}
where
$4p_1\cdot k_1\approx 4E_1\omega_1(1-\cos\theta_1)=8E_1\omega_1\cos^2(\vartheta/2)$.
To facilitate comparison with the vortex case, we obtain the differential cross section in $(\omega_2,\theta_2)$. Although there is generally no axial symmetry for scattering with definite helicities, we can choose, without loss of generality, the initial photon to be in the $xz$ plane so that $\varphi_1=0$ effectively. Recalling the definitions
\begin{subequations}
  \begin{align}
    &\pvec_1=p_1(0,0,1),\quad
    \kvec_1=\omega_1(\sin\theta_1,0,\cos\theta_1),
    \\
    &\pvec_2=p_2(\sin\theta_e\cos\varphi_e,\sin\theta_e\sin\varphi_e,\cos\theta_e),\quad
    \kvec_2=\omega_2(\sin\theta_2\cos\varphi_2,\sin\theta_2\sin\varphi_2,\cos\theta_2),
  \end{align}
\end{subequations}
where $\theta_1=\pi-\vartheta$ with $0\leq\vartheta<\pi/2$, the $p_2$ and $\theta_e$ integrals are completed by the $\delta$ functions of the energy and longitudinal momentum and $(E_2,p_2,\theta_e)$ are uniquely determined in terms of $(p_a,\omega_a,\theta_a)$ 
\begin{subequations}
    \begin{align}
&E_2=E_1+\omega_1-\omega_2,\quad 
p_2=\sqrt{E_2^2-M^2},
\\
&\cos\theta_e=p_2^{-1}[p_1-\omega_1\cos\vartheta-\omega_2\cos\theta_2],\quad
p_2^z=p_2\cos\theta_e,\quad
\kappa_e=p_2\sin\theta_e.
    \end{align}
\end{subequations}
The $\varphi_e$ and $\varphi_2$ integrals are done in \cref{eq:planar_integral} but with $\varphi_1=0$.
In summary, indicating the $\varphi_{2,e}$ dependence in $\calA_\pw(\varphi_2,\varphi_e)$, the required differential cross section is
\begin{align}
  \frac{d\sigma}{\omega_2d\omega_2 d\cos\theta_2}
  =\frac{1}{4p_1\cdot k_1}
  \frac{1}{(4\pi)^2}\frac{1}{2\Delta}
  \sum_{\eta=\pm}|\calA_\pw(-\eta\delta_2,\eta\delta_e)|^2.
\end{align}
The singularity in $\Delta^{-1}$ is integrable, different from the case of Bessel-vortex scattering.

\twocolumngrid
\bibliography{references.bib}{}

@book{torres2011twisted,
  title={Twisted photons: applications of light with orbital angular momentum},
  author={Torres, Juan P and Torner, Lluis},
  year={2011},
  publisher={John Wiley \& Sons}
}

@book{andrews2012angular,
  title={The angular momentum of light},
  author={Andrews, David L and Babiker, Mohamed},
  year={2012},
  publisher={Cambridge University Press}
}

@incollection{ALLEN1999291,
title = {IV The Orbital Angular Momentum of Light},
editor = {E. Wolf},
series = {Progress in Optics},
publisher = {Elsevier},
volume = {39},
pages = {291-372},
year = {1999},
issn = {0079-6638},
doi = {https://doi.org/10.1016/S0079-6638(08)70391-3},
url = {https://www.sciencedirect.com/science/article/pii/S0079663808703913},
author = {L. Allen and M.J. Padgett and M. Babiker}
}

@article{Molina-Terriza:2007ydx,
    author = "Molina-Terriza, Gabriel and Torres, Juan P. and Torner, Lluis",
    title = "{Twisted photons}",
    doi = "10.1038/nphys607",
    journal = "Nature Phys.",
    volume = "3",
    number = "5",
    pages = "305--310",
    year = "2007"
}

@article{Padgett:17,
author = {Miles J. Padgett},
journal = {Opt. Express},
number = {10},
pages = {11265--11274},
publisher = {Optica Publishing Group},
title = {Orbital angular momentum 25 years on [Invited]},
volume = {25},
month = {May},
year = {2017},
url = {https://opg.optica.org/oe/abstract.cfm?URI=oe-25-10-11265},
doi = {10.1364/OE.25.011265}
}

@article{Knyazev:2018,
	author = {B. A. Knyazev and V. G. Serbo},
	title = {Beams of photons with nonzero orbital angular momentum projection: new results},
	publisher = {Physics-Uspekhi},
	year = {2018},
	journal = {Phys. Usp.},
	volume = {61},
	number = {5},
	pages = {449-479},
	url = {https://ufn.ru/en/articles/2018/5/e/},
	doi = {10.3367/UFNe.2018.02.038306}
}

@article{Wu:2021trm,
    author = "Wu, Yuanbin and Gargiulo, Simone and Carbone, Fabrizio and Keitel, Christoph H. and P{\'a}lffy, Adriana",
    title = "{Dynamical Control of Nuclear Isomer Depletion via Electron Vortex Beams}",
    eprint = "2107.12448",
    archivePrefix = "arXiv",
    primaryClass = "physics.atom-ph",
    doi = "10.1103/PhysRevLett.128.162501",
    journal = "Phys. Rev. Lett.",
    volume = "128",
    number = "16",
    pages = "162501",
    year = "2022"
}

@article{Lu:2023wrf,
    author = "Lu, Zhi-Wei and others",
    title = "{Manipulation of Giant Multipole Resonances via Vortex {\ensuremath{\gamma}} Photons}",
    eprint = "2306.08377",
    archivePrefix = "arXiv",
    primaryClass = "nucl-th",
    doi = "10.1103/PhysRevLett.131.202502",
    journal = "Phys. Rev. Lett.",
    volume = "131",
    number = "20",
    pages = "202502",
    year = "2023"
}

@article{Ivanov:2019vxe,
    author = "Ivanov, Igor P. and Korchagin, Nikolai and Pimikov, Alexandr and Zhang, Pengming",
    title = "{Doing spin physics with unpolarized particles}",
    eprint = "1911.08423",
    archivePrefix = "arXiv",
    primaryClass = "hep-ph",
    reportNumber = "CFTP/19-030",
    doi = "10.1103/PhysRevLett.124.192001",
    journal = "Phys. Rev. Lett.",
    volume = "124",
    number = "19",
    pages = "192001",
    year = "2020"
}

@article{Ivanov:2020kcy,
    author = "Ivanov, Igor P. and Korchagin, Nikolai and Pimikov, Alexandr and Zhang, Pengming",
    title = "{Twisted particle collisions: a new tool for spin physics}",
    eprint = "2002.01703",
    archivePrefix = "arXiv",
    primaryClass = "hep-ph",
    reportNumber = "CFTP/20-001",
    doi = "10.1103/PhysRevD.101.096010",
    journal = "Phys. Rev. D",
    volume = "101",
    number = "9",
    pages = "096010",
    year = "2020"
}

@article{Chen_2018,
   title={Gamma-Ray Beams with Large Orbital Angular Momentum via Nonlinear Compton Scattering with Radiation Reaction},
   volume={121},
   ISSN={1079-7114},
   url={http://dx.doi.org/10.1103/PhysRevLett.121.074801},
   DOI={10.1103/physrevlett.121.074801},
   number={7},
   journal={Physical Review Letters},
   publisher={American Physical Society (APS)},
   author={Chen, Yue-Yue and Li, Jian-Xing and Hatsagortsyan, Karen Z. and Keitel, Christoph H.},
   year={2018},
   month=Aug }

@article{Lu:2025tyx,
    author = "Lu, Zhi-Wei and Zhang, Hanxu and Li, Tao and Ababekri, Mamutjan and Wang, Xu and Li, Jian-Xing",
    title = "{Nuclear excitation and control induced by intense vortex lasers}",
    eprint = "2503.12812",
    archivePrefix = "arXiv",
    primaryClass = "nucl-th",
    doi = "10.1103/6m8k-9pcb",
    journal = "Phys. Rev. C",
    volume = "113",
    number = "4",
    pages = "044315",
    year = "2026"
}

@article{Babiker_2019,
doi = {10.1088/2040-8986/aaed14},
url = {https://doi.org/10.1088/2040-8986/aaed14},
year = {2018},
month = {dec},
publisher = {IOP Publishing},
volume = {21},
number = {1},
pages = {013001},
author = {Babiker, Mohamed and Andrews, David L and Lembessis, Vassilis E},
title = {Atoms in complex twisted light},
journal = {Journal of Optics}
}

@article{Bliokh:2017uvr,
    author = "Bliokh, K. Y. and others",
    title = "{Theory and applications of free-electron vortex states}",
    eprint = "1703.06879",
    archivePrefix = "arXiv",
    primaryClass = "quant-ph",
    doi = "10.1016/j.physrep.2017.05.006",
    journal = "Phys. Rept.",
    volume = "690",
    pages = "1--70",
    year = "2017"
}

@article{Lloyd:2017ipi,
    author = "Lloyd, S. {\,}M. and Babiker, M. and Thirunavukkarasu, G. and Yuan, J.",
    title = "{Electron vortices: Beams with orbital angular momentum}",
    doi = "10.1103/RevModPhys.89.035004",
    journal = "Rev. Mod. Phys.",
    volume = "89",
    number = "3",
    pages = "035004",
    year = "2017"
}

@article{Knyazev_2018,
    author = {Knyazev, B A and Serbo, V G},
    title = {Beams of photons with nonzero projections of orbital angular momenta: new results},
    doi = {10.3367/UFNe.2018.02.038306},
    journal = {Physics-Uspekhi},
    volume = {61},
    number = {5},
    pages = {449},
    year = {2018},
    month = {may},
    publisher = {Uspekhi Fizicheskikh Nauk, Russian Academy of Sciences},
}

@article{Ivanov:2022jzh,
    author = "Ivanov, Igor P.",
    title = "{Promises and challenges of high-energy vortex states collisions}",
    eprint = "2205.00412",
    archivePrefix = "arXiv",
    primaryClass = "hep-ph",
    doi = "10.1016/j.ppnp.2022.103987",
    journal = "Prog. Part. Nucl. Phys.",
    volume = "127",
    pages = "103987",
    year = "2022"
}

@article{Allen:1992zz,
    author = "Allen, L. and Beijersbergen, M. W. and Spreeuw, R. J. C. and Woerdman, J. P.",
    title = "{Orbital angular momentum of light and the transformation of Laguerre-Gaussian laser modes}",
    doi = "10.1103/PhysRevA.45.8185",
    journal = "Phys. Rev. A",
    volume = "45",
    pages = "8185--8189",
    year = "1992"
}

@article{Ivanov:2011tu,
  author        = {Ivanov, Igor P.},
  title         = {{Creation of two vortex-entangled beams in a vortex beam collision with a plane wave}},
  eprint        = {1110.5760},
  archiveprefix = {arXiv},
  primaryclass  = {quant-ph},
  doi           = {10.1103/PhysRevA.85.033813},
  journal       = {Phys. Rev. A},
  volume        = {85},
  pages         = {033813},
  year          = {2012}
}

@article{Jentschura:2010ap,
    author = "Jentschura, U. D. and Serbo, V. G.",
    title = "{Generation of High-Energy Photons with Large Orbital Angular Momentum by Compton Backscattering}",
    eprint = "1008.4788",
    archivePrefix = "arXiv",
    primaryClass = "physics.acc-ph",
    doi = "10.1103/PhysRevLett.106.013001",
    journal = "Phys. Rev. Lett.",
    volume = "106",
    number = "1",
    pages = "013001",
    year = "2011"
}

@article{Jentschura:2011ih,
  author        = {Jentschura, U. D. and Serbo, V. G.},
  title         = {{Compton Upconversion of Twisted Photons: Backscattering of Particles with Non-Planar Wave Functions}},
  eprint        = {1101.1206},
  archiveprefix = {arXiv},
  primaryclass  = {physics.acc-ph},
  doi           = {10.1140/epjc/s10052-011-1571-z},
  journal       = {Eur. Phys. J. C},
  volume        = {71},
  pages         = {1571},
  year          = {2011}
}

@article{Ivanov:2011bv,
    author = "Ivanov, I. P. and Serbo, V. G.",
    title = "{Scattering of Twisted Particles: Extension to Wave Packets and orbital helicity}",
    eprint = "1105.6244",
    archivePrefix = "arXiv",
    primaryClass = "hep-ph",
    doi = "10.1103/PhysRevA.84.033804",
    journal = "Phys. Rev. A",
    volume = "84",
    pages = "033804",
    year = "2011"
}

@article{Seipt:2014bxa,
    author = "Seipt, D. and Surzhykov, A. and Fritzsche, S.",
    title = "{Structured x-ray beams from twisted electrons by inverse Compton scattering of laser light}",
    eprint = "1407.4329",
    archivePrefix = "arXiv",
    primaryClass = "physics.optics",
    doi = "10.1103/PhysRevA.90.012118",
    journal = "Phys. Rev. A",
    volume = "90",
    number = "1",
    pages = "012118",
    year = "2014"
}

@article{Ivanov:2011kk,
    author = "Ivanov, I. P.",
    title = "{Colliding particles carrying non-zero orbital angular momentum}",
    eprint = "1101.5575",
    archivePrefix = "arXiv",
    primaryClass = "hep-ph",
    doi = "10.1103/PhysRevD.83.093001",
    journal = "Phys. Rev. D",
    volume = "83",
    pages = "093001",
    year = "2011"
}

@article{Ivanov:2012na,
    author = "Ivanov, I. P.",
    title = "{Measuring the phase of the scattering amplitude with vortex beams}",
    eprint = "1201.5040",
    archivePrefix = "arXiv",
    primaryClass = "hep-ph",
    doi = "10.1103/PhysRevD.85.076001",
    journal = "Phys. Rev. D",
    volume = "85",
    pages = "076001",
    year = "2012"
}

@article{Ivanov:2016oue,
    author = "Ivanov, I. P. and Seipt, D. and Surzhykov, A. and Fritzsche, S.",
    title = "{Elastic scattering of vortex electrons provides direct access to the Coulomb phase}",
    eprint = "1608.06551",
    archivePrefix = "arXiv",
    primaryClass = "hep-ph",
    doi = "10.1103/PhysRevD.94.076001",
    journal = "Phys. Rev. D",
    volume = "94",
    number = "7",
    pages = "076001",
    year = "2016"
}

@article{Karlovets:2016dva,
    author = "Karlovets, Dmitry V.",
    title = "{Probing phase of a scattering amplitude beyond the plane-wave approximation}",
    eprint = "1608.08858",
    archivePrefix = "arXiv",
    primaryClass = "hep-ph",
    doi = "10.1209/0295-5075/116/31001",
    journal = "EPL",
    volume = "116",
    pages = "31001",
    year = "2016"
}

@article{Karlovets:2016jrd,
    author = "Karlovets, Dmitry",
    title = "{Scattering of wave packets with phases}",
    eprint = "1611.08302",
    archivePrefix = "arXiv",
    primaryClass = "hep-ph",
    doi = "10.1007/JHEP03(2017)049",
    journal = "JHEP",
    volume = "03",
    pages = "049",
    year = "2017"
}

@article{Korchagin:2024nen,
    author = "Korchagin, Nikolai",
    title = "{Studying timelike proton form factors using vortex state pp{\textasciimacron} annihilation}",
    eprint = "2403.08949",
    archivePrefix = "arXiv",
    primaryClass = "hep-ph",
    doi = "10.1103/PhysRevD.111.076005",
    journal = "Phys. Rev. D",
    volume = "111",
    number = "7",
    pages = "076005",
    year = "2025"
}

@article{Zhao:2023cwd,
    author = "Zhao, Pengcheng",
    title = "{Momentum space oscillation properties of vortex states collision}",
    eprint = "2312.00424",
    archivePrefix = "arXiv",
    primaryClass = "hep-ph",
    doi = "10.1103/PhysRevD.109.096015",
    journal = "Phys. Rev. D",
    volume = "109",
    number = "9",
    pages = "096015",
    year = "2024"
}

@article{Yang:2026byv,
    author = "Yang, Yaoqi and Ivanov, Igor P.",
    title = "{Universal features of high-energy scattering of Laguerre-Gaussian states}",
    eprint = "2604.00575",
    archivePrefix = "arXiv",
    primaryClass = "hep-ph",
    doi = "10.1103/h5bv-6sl5",
    journal = "Phys. Rev. D",
    volume = "113",
    number = "11",
    pages = "116020",
    year = "2026"
}

@article{Liao:2026gqh,
    author = "Liao, Yi and Wang, Quan-Yu and Wu, Yuanbin",
    title = "{Off-axis vortex scattering of electron-positron annihilation into a photon pair}",
    eprint = "2601.07530",
    archivePrefix = "arXiv",
    primaryClass = "hep-ph",
    month = "1",
    year = "2026"
}

@article{Wei:2025zsv,
    author = "Wei, Mingxuan and others",
    title = "{Experimental Evidence of Vortex {\ensuremath{\gamma}} Photons in All-Optical Inverse Compton Scattering}",
    eprint = "2503.18843",
    archivePrefix = "arXiv",
    primaryClass = "physics.plasm-ph",
    doi = "10.1103/92v4-bzp2",
    journal = "Phys. Rev. Lett.",
    volume = "136",
    number = "2",
    pages = "025001",
    year = "2026"
}

@article{Heckenberg1992,
  author  = {Heckenberg, N. R. and McDuff, R. and Smith, C. P. and White, A. G.},
  title   = {Generation of optical phase singularities by computer-generated holograms},
  journal = {Optics Letters},
  volume  = {17},
  number  = {3},
  pages   = {221--223},
  year    = {1992},
  doi     = {10.1364/OL.17.000221}
}

@article{Sueda2004,
  author  = {Sueda, K. and Miyaji, G. and Miyanaga, N. and Nakatsuka, M.},
  title   = {Laguerre-Gaussian beam generated with a multilevel spiral phase plate for high intensity laser pulses},
  journal = {Optics Express},
  volume  = {12},
  number  = {15},
  pages   = {3548--3553},
  year    = {2004},
  doi     = {10.1364/OPEX.12.003548}
}

@article{YaoPadgett2011,
  author  = {Yao, Alison M. and Padgett, Miles J.},
  title   = {Orbital angular momentum: Origins, behavior and applications},
  journal = {Advances in Optics and Photonics},
  volume  = {3},
  number  = {2},
  pages   = {161--204},
  year    = {2011},
  doi     = {10.1364/AOP.3.000161}
}

@article{Willner2015,
  author  = {Willner, Alan E. and Huang, H. and Yan, Y. and Ren, Y. and Ahmed, N. and Xie, G. and Bao, C. and Li, L. and Cao, Y. and Zhao, Z. and Wang, J. and Lavery, Martin P. J. and Tur, Moshe and Ramachandran, S. and Molisch, A. F. and Ashrafi, N. and Ashrafi, S.},
  title   = {Optical communications using orbital angular momentum beams},
  journal = {Advances in Optics and Photonics},
  volume  = {7},
  number  = {1},
  pages   = {66--106},
  year    = {2015},
  doi     = {10.1364/AOP.7.000066}
}

@article{Mair2001,
  author  = {Mair, Alois and Vaziri, Alipasha and Weihs, Gregor and Zeilinger, Anton},
  title   = {Entanglement of the orbital angular momentum states of photons},
  journal = {Nature},
  volume  = {412},
  number  = {6844},
  pages   = {313--316},
  year    = {2001},
  doi     = {10.1038/35085529}
}

@article{He1995,
  author  = {He, H. and Friese, M. E. J. and Heckenberg, N. R. and Rubinsztein-Dunlop, H.},
  title   = {Direct observation of transfer of angular momentum to absorptive particles from a laser beam with a phase singularity},
  journal = {Physical Review Letters},
  volume  = {75},
  number  = {5},
  pages   = {826--829},
  year    = {1995},
  doi     = {10.1103/PhysRevLett.75.826}
}

@article{Peele2002,
  author  = {Peele, Andrew G. and McMahon, Philip J. and Paterson, David and Tran, Chanh Q. and Mancuso, Adrian P. and Nugent, Keith A. and Hayes, Jason P. and Harvey, Erol and Lai, Barry and McNulty, Ian},
  title   = {Observation of an x-ray vortex},
  journal = {Optics Letters},
  volume  = {27},
  number  = {20},
  pages   = {1752--1754},
  year    = {2002},
  doi     = {10.1364/OL.27.001752}
}

@article{Bahrdt2013,
  author  = {Bahrdt, Johannes and Holldack, Karsten and Kuske, Peter and M{\"u}ller, Roland and Scheer, Michael and Schmid, Peter},
  title   = {First observation of photons carrying orbital angular momentum in undulator radiation},
  journal = {Physical Review Letters},
  volume  = {111},
  number  = {3},
  pages   = {034801},
  year    = {2013},
  doi     = {10.1103/PhysRevLett.111.034801}
}

@article{Gauthier2017,
  author  = {Gauthier, David and Ribi{\v{c}}, Primo{\v{z}} Rebernik and Adhikary, G. and Camper, A. and Chappuis, C. and Cucini, R. and DiMauro, L. F. and Dovillaire, G. and Frassetto, F. and G{\'e}neaux, R. and Miotti, P. and Poletto, L. and Ressel, B. and Spezzani, C. and Stupar, M. and Ruchon, T. and De Ninno, G.},
  title   = {Tunable orbital angular momentum in high-harmonic generation},
  journal = {Nature Communications},
  volume  = {8},
  pages   = {14971},
  year    = {2017},
  doi     = {10.1038/ncomms14971}
}

@article{Ribic2017,
  author  = {Ribi{\v{c}}, Primo{\v{z}} Rebernik and R{\"o}sner, Benedikt and Gauthier, David and others},
  title   = {Extreme-ultraviolet vortices from a free-electron laser},
  journal = {Physical Review X},
  volume  = {7},
  number  = {3},
  pages   = {031036},
  year    = {2017},
  doi     = {10.1103/PhysRevX.7.031036}
}

@article{Lee2019,
  author  = {Lee, J. C. T. and Alexander, S. J. and Kevan, S. D. and Roy, S. and McMorran, B. J.},
  title   = {Laguerre--Gauss and Hermite--Gauss soft x-ray states generated using diffractive optics},
  journal = {Nature Photonics},
  volume  = {13},
  number  = {3},
  pages   = {205--209},
  year    = {2019},
  doi     = {10.1038/s41566-018-0328-8}
}

@article{Uchida:2010hbm,
    author = "Uchida, Masaya and Tonomura, Akira",
    title = "{Generation of electron beams carrying orbital angular momentum}",
    doi = "10.1038/nature08904",
    journal = "Nature",
    volume = "464",
    number = "7289",
    pages = "737--739",
    year = "2010"
}

@article{Verbeeck:2010ezk,
    author = "Verbeeck, J. and Tian, H. and Schattschneider, P.",
    title = "{Production and application of electron vortex beams}",
    doi = "10.1038/nature09366",
    journal = "Nature",
    volume = "467",
    number = "7313",
    pages = "301--304",
    year = "2010"
}

@article{McMorran:2011bql,
    author = "McMorran, Benjamin J. and Agrawal, Amit and Anderson, Ian M. and Herzing, Andrew A. and Lezec, Henri J. and McClelland, Jabez J. and Unguris, John",
    title = "{Electron Vortex Beams with High Quanta of Orbital Angular Momentum}",
    doi = "10.1126/science.1198804",
    journal = "Science",
    volume = "331",
    number = "6014",
    pages = "1198804",
    year = "2011"
}

@article{Karlovets:2022evc,
    author = "Karlovets, D. V. and Baturin, S. S. and Geloni, G. and Sizykh, G. K. and Serbo, V. G.",
    title = "{Generation of vortex particles via generalized measurements}",
    eprint = "2201.07997",
    archivePrefix = "arXiv",
    primaryClass = "hep-ph",
    doi = "10.1140/epjc/s10052-022-10991-w",
    journal = "Eur. Phys. J. C",
    volume = "82",
    number = "11",
    pages = "1008",
    year = "2022"
}

@article{Karlovets:2022mhb,
    author = "Karlovets, D. V. and Baturin, S. S. and Geloni, G. and Sizykh, G. K. and Serbo, V. G.",
    title = "{Shifting physics of vortex particles to higher energies via quantum entanglement}",
    eprint = "2203.12012",
    archivePrefix = "arXiv",
    primaryClass = "hep-ph",
    doi = "10.1140/epjc/s10052-023-11529-4",
    journal = "Eur. Phys. J. C",
    volume = "83",
    number = "5",
    pages = "372",
    year = "2023"
}

@article{Liao:2025skb,
    author = "Liao, Yi and Wang, Quan-Yu and Wu, Yuanbin",
    title = "{All-vortex nonlinear Compton scattering in a polarized laser field}",
    eprint = "2504.04425",
    archivePrefix = "arXiv",
    primaryClass = "hep-ph",
    doi = "10.1103/gxs8-vgbj",
    journal = "Phys. Rev. D",
    volume = "112",
    number = "3",
    pages = "033004",
    year = "2025"
}
\bibliographystyle{utphys}

\end{document}